\documentclass[aps,prl,graphicx,9pt,twocolumn,superscriptaddress]{revtex4-2}
\usepackage{amsmath,amscd,graphicx,graphicx,cases}
\usepackage{multirow,float,subfigure,appendix}
\usepackage{diagbox,makecell,blindtext,color,amssymb}
\usepackage[colorlinks,linkcolor=blue,urlcolor=blue,
anchorcolor=blue,citecolor=blue]{hyperref}
\usepackage{textcomp,booktabs,colortbl,easyReview}
\newcolumntype{C}{c<{\kern\tabcolsep}@{}}
\usepackage{times}
\usepackage{lineno}
\usepackage{xcolor,ulem}

\begin{document}

\title{Observation of a superradiant phase transition with emergent cat states} 

\author{Ri-Hua Zheng}
\thanks{These authors contribute equally to this work.}
\author{Wen Ning}
\thanks{These authors contribute equally to this work.}
\affiliation{Fujian Key Laboratory of Quantum Information and Quantum
	Optics, College of Physics and Information Engineering, Fuzhou University,
	Fuzhou, Fujian 350108, China}
\author{Ye-Hong Chen}
\thanks{These authors contribute equally to this work.}
\affiliation{Fujian Key Laboratory of Quantum Information and Quantum
	Optics, College of Physics and Information Engineering, Fuzhou University,
	Fuzhou, Fujian 350108, China}
\affiliation{Theoretical Quantum Physics Laboratory, RIKEN Cluster for Pioneering 
	Research, Wako-shi, Saitama 351-0198, Japan}
\affiliation{Quantum Information Physics Theory Research Team, RIKEN Center for Quantum Computing (RQC), Wako-shi, Saitama 351-0198, Japan}

\author{Jia-Hao L\"{u}} 
\affiliation{Fujian Key Laboratory of Quantum Information and Quantum
	Optics, College of Physics and Information Engineering, Fuzhou University,
	Fuzhou, Fujian 350108, China}

\author{Li-Tuo Shen} 
\affiliation{Fujian Key Laboratory of Quantum Information and Quantum
	Optics, College of Physics and Information Engineering, Fuzhou University,
	Fuzhou, Fujian 350108, China}

\author{Kai Xu}
\affiliation{Institute of Physics and Beijing National Laboratory for Condensed
	Matter Physics, Chinese Academy of Sciences, Beijing 100190, China}
\affiliation{CAS Center for Excellence in Topological Quantum Computation,
	University of Chinese Academy of Sciences, Beijing 100190, China}

\author{Yu-Ran Zhang}
\affiliation{School of Physics and Optoelectronics, South China University of Technology, Guangzhou 510640, China}
\affiliation{Theoretical Quantum Physics Laboratory, RIKEN Cluster for Pioneering Research, Wako-shi, Saitama 351-0198, Japan}
\affiliation{Quantum Information Physics Theory Research Team, RIKEN Center for Quantum Computing (RQC), Wako-shi, Saitama 351-0198, Japan}

\author{Da Xu}
\affiliation{Interdisciplinary Center of Quantum Information, State Key Laboratory of Modern Optical Instrumentation
	and Zhejiang Province Key Laboratory of Quantum Technology and Device, School of Physics, Zhejiang University,
	Hangzhou 310027, China}

\author{Hekang Li}
\affiliation{CAS Center for Excellence in Topological Quantum Computation,
	University of Chinese Academy of Sciences, Beijing 100190, China}


\author{Yan Xia}
\affiliation{Fujian Key Laboratory of Quantum Information and Quantum
	Optics, College of Physics and Information Engineering, Fuzhou University,
	Fuzhou, Fujian 350108, China}

\author{Fan Wu}
\affiliation{Fujian Key Laboratory of Quantum Information and Quantum
	Optics, College of Physics and Information Engineering, Fuzhou University,
	Fuzhou, Fujian 350108, China}

\author{Zhen-Biao Yang}\email{zbyang@fzu.edu.cn}
\affiliation{Fujian Key Laboratory of Quantum Information and Quantum
	Optics, College of Physics and Information Engineering, Fuzhou University,
	Fuzhou, Fujian 350108, China}

\author{Adam Miranowicz}
\affiliation{Theoretical Quantum Physics Laboratory, RIKEN Cluster for Pioneering Research, Wako-shi, Saitama 351-0198, Japan}
\affiliation{Institute of Spintronics and Quantum Information, Faculty of Physics, Adam Mickiewicz University, 61-614 Pozna\'{n}, Poland}

\author{Neill Lambert}
\affiliation{Theoretical Quantum Physics Laboratory, RIKEN Cluster for Pioneering 
	Research, Wako-shi, Saitama 351-0198, Japan}

\author{Dongning Zheng}
\affiliation{Institute of Physics and Beijing National Laboratory for Condensed
	Matter Physics, Chinese Academy of Sciences, Beijing 100190, China}
\affiliation{CAS Center for Excellence in Topological Quantum Computation,
	University of Chinese Academy of Sciences, Beijing 100190, China}

\author{Heng Fan}
\affiliation{Institute of Physics and Beijing National Laboratory for Condensed
	Matter Physics, Chinese Academy of Sciences, Beijing 100190, China}
\affiliation{CAS Center for Excellence in Topological Quantum Computation,
	University of Chinese Academy of Sciences, Beijing 100190, China}

\author{Franco 	Nori}\email{fnori@riken.jp}
\affiliation{Theoretical Quantum Physics Laboratory, RIKEN Cluster for Pioneering 
	Research, Wako-shi, Saitama 351-0198, Japan}
\affiliation{Quantum Information Physics Theory Research Team, RIKEN Center for Quantum Computing (RQC), Wako-shi, Saitama 351-0198, Japan}
\affiliation{Department of Physics, University of Michigan, Ann Arbor, Michigan
	48109-1040, USA}

\author{ Shi-Biao Zheng}\email{t96034@fzu.edu.cn}
\affiliation{Fujian Key Laboratory of Quantum Information and Quantum
	Optics, College of Physics and Information Engineering, Fuzhou University,
	Fuzhou, Fujian 350108, China}

\begin{abstract}
Superradiant phase transitions (SPTs) are important for understanding light-matter
interactions at the quantum level, and play a central role in
criticality-enhanced quantum sensing. So far, SPTs have been observed in driven-dissipative systems, but the emergent light fields did not show any nonclassical characteristic due to the presence of strong dissipation. Here we report an experimental demonstration of the SPT featuring the emergence of a highly nonclassical photonic field, realized with a resonator coupled to a superconducting qubit, implementing the quantum Rabi model. We fully characterize the light-matter state by Wigner matrix tomography. The measured matrix elements exhibit quantum interference intrinsic of a photonic mesoscopic superposition, and reveal light-matter entanglement. 
\end{abstract}

\maketitle 

%

The Dicke model \cite{Dicke1954Pr,Kirton2019}, involving a quantized light field coupled to $N$ two-level atoms, represents a paradigm for realizing exotic quantum phenomena that are absent in semiclassical light-matter systems. Superradiant phase transitions (SPTs) are one of the most famous examples \cite{Wang1973,Hepp1973,Lambert2004}, where the behavior of the light is sharply changed when the light-matter coupling strength becomes comparable to their frequencies.
Under {equilibrium conditions}, the SPT features a sudden buildup of a photonic field that is highly entangled with the atoms in a mesoscopic superposition \cite{Lambert2004}. In addition to its fundamental appeal, such cat states can be used as an
intrinsically-protected qubit for fault-tolerant quantum computation \cite{Nataf2011} and as a resource for quantum enhanced metrology \cite{Kwon2019}. The equilibrium SPT has been attracting enduring attention since the 1970s, but its experimental demonstration
still remains very challenging. This is mainly because the neglected square of the vector potential actually increases quadratically with the coupling strength and the photon number, which prohibits
the occurrence of SPTs, known as the no-go theorem \cite{Rzazewski1975}. Over the past decade, breakthrough experiments have been reported for dynamical realizations of SPTs with a collection of driven atoms trapped in an optical cavity \cite{Zhang2017,Baumann2010,Baumann2011,Brennecke2013,Klinder2015,Leonard2017,Ferri2021,Zhang2021}, whose photonic dissipation enabled the phase transition to be monitored by measuring the
output field. This dissipation, however, at the same time obscured the quantum coherence of the light, as well as the light-matter entanglement inherent in the superradiant phase (SP).

Although originally proposed in the thermodynamic limit $N\rightarrow
\infty $, SPTs can actually occur in the quantum Rabi model (QRM), which only involves a
single atom coupled to a light field \cite{Ashhab2013,Hwang2015}. Recent years have witnessed
remarkable advances in simulations of the Rabi model in different systems,
where the photonic mode was emulated by a phononic mode of a trapped ion \cite{Lv2018,Cai2021},  
{while the light field coupling the ion's internal and external degrees of freedom} is classical.
Circuit QED represents an
alternative excellent platform for exploring quantized light-matter systems in regimes that are inaccessible 
with conventional cavity QED \cite{Kockum2019,FornDiaz2019,Gu2017,Ballester2012},
and for simulating controlled many-body dynamics \cite{Feng2015,XuKaiSCIADV}. 
In
particular, recent experiments \cite{Yoshihara2018} have demonstrated some spectroscopic
signatures in the deep-strong coupling regimes. 
The long coherence times of the superconducting qubits and the microwave photons makes circuit QED promising for realizing SPT produced by a unitary process, in distinct contrast with ultracold-atoms-based cavity QED systems \cite{Zhang2017,Baumann2010,Baumann2011,Brennecke2013,Klinder2015,Leonard2017,Ferri2021,Zhang2021}, where SPT was realized in a dissipative-driven manner. This unitary nature, together with the ability to individually control and measure the superconducting qubits, enables the exploration of nonclassical characteristics associated with the SPT, such as the qubit-resonator entanglement and phase-space quantum interference behaviors of the resonator.

Theoretical investigations indicate that
the no-go theorem can be circumvented in circuit QED 
systems \cite{Nataf2010}. However,
the approximation for describing a superconducting artificial atom as a
qubit may break down when increasing the coupling strength due to the
limited anharmonicity \cite{Bernardis2018}, which prohibits the occurrence of dynamical signatures of the Rabi model,
even when the deep-strong regime is reached \cite{Yoshihara2018}. To
overcome this problem, it was proposed to effectively transform the Jaynes-Cummings model into the Rabi model, by applying continuous 
microwave fields to the qubit \cite{FornDiaz2019,Sanchez2020}
or by introducing a two-photon drive to the resonator \cite{Qin2018,Leroux2018,Zhu2020,ChenPrl2021}. 
Following these approaches, some important features predicted  {by} the Rabi model have been observed \cite{Langford2017,Braumller2017}. 
Despite these advancements, so far the SPT of a real radiant field with nonclassical features has not been reported in any system.

Here we report {a realization of the first-order SPT} of a quantum
light field manifested by an emergent cat state. Our demonstration involves a resonator and a
superconducting qubit coupled at the second sideband of a strong parametric
modulation produced by an ac magnetic flux. This strong longitudinal modulation, together with a weak modulation and
a transverse microwave driving, enables the realization of an effective Rabi model with a controllable coupling-frequency ratio. 
We fully describe the 
nonclassical behavior of the system by measuring the Wigner function matrix of the joint
qubit-resonator system, which contains full information about its 
state. The measured matrix elements unambiguously demonstrate that the photonic field emergent in the SP is in a quantum superposition of 
two quasi-classical states that are degenerate in amplitude but have opposite phases. These results bridge the gap between the phase transitions predicted in 
closed quantum systems and those observed in real macroscopic systems, which is critical to understanding how a symmetry-broken macroscopic order emerges 
from the dynamics governed by a symmetry-preserving Hamiltonian.


\begin{figure*}
	\centering
	\includegraphics[width=18cm]{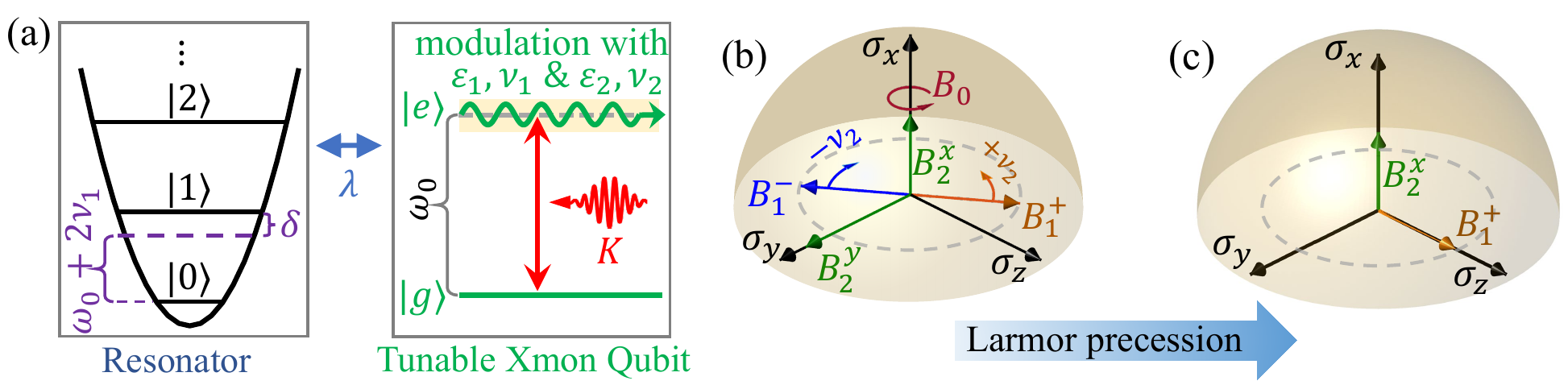}
	\caption{Theoretical model. (a)
		Sketch for the qubit-resonator coupling. The test qubit is coupled to the
		resonator at the second sideband of a sine longitudinal modulation with modulating
		amplitude $\varepsilon _{1}$ and frequency $\nu _{1}$.
		A second sine modulation with amplitude $\varepsilon _{2}$ and frequency $\nu _{2}$ is used to control the effective frequency of the qubit. 
		These two modulations, together with a transverse drive $K$, effectively realizes an effective Rabi Hamiltonian.
		Bloch representations in (b) the laboratory frame and (c) the precessing frame. By analogy with the motion of a
		spin-1/2, the transverse drive can be regarded as a static magnetic
		field of strength $B_{0}\propto K$ along the $x$-axis, which forces the 
		Bloch vector of the qubit to precess with angular frequency $B_{0}$. The second
		longitudinal modulation corresponds to applying two magnetic fields on the $yz$-plane with the same amplitude $|B_{1}^{\pm}|=\varepsilon_{2}/4$, rotating at the same angular frequency $\nu_{2}=B_{0}$, but in opposite directions. The light field stored in the resonator acts as an effective magnetic field with components $B_{2}^{x}$ and $B_{2}^{y}$. In the precessing frame associated with $B_0$, the components $B_{1}^{+}$ and $B_{2}^{x}$ are aligned with the $z$- and $x$-axes, respectively. The remaining components have negligible effects in the rotating-wave approximation (not shown).
	}
	\label{fig1}
\end{figure*}

The theoretical model includes a quantized light field stored in a
resonator coupled to a qubit, e.g., a tunable Xmon qubit [see Fig.~1(a)], whose transition frequency is periodically
modulated \cite{Wang2019} as
$\omega _{q}=\omega _{0}+\varepsilon _{1}\cos (\nu_{1}t)+\varepsilon _{2}\cos (\nu _{2}t)$, 
where $\omega _{0}$ corresponds to the mean transition frequency of the
qubit, and $\varepsilon _{1,(2)}$ [$\nu _{1,(2)}$] are the corresponding
modulation amplitudes (frequencies). In addition to these
longitudinal modulations, the qubit is transversely driven by an external
field at the frequency $\omega _{0}$ with an amplitude $K$. The system dynamics is described by the
Hamiltonian%
\begin{eqnarray}
	H =\hbar \left[\omega _{p}a^{\dagger }a+\frac{ \omega _{q}}{2}\sigma_{z} 
	+\left(\lambda a^{\dagger }\sigma_{-} +Ke^{i\omega _{0}t} \sigma_{-} +{\rm{h.c.}}\right)\right], \cr
\end{eqnarray}%
where $a^{\dagger }$ ($a$) denotes the creation (annihilation) operator
for the photonic field with frequency $\omega _{p}$, $\lambda $ is the
qubit-resonator coupling strength, $\sigma_{z}=|e\rangle\langle e|-|g\rangle\langle g|$, and $\sigma_{-}=|g\rangle\langle e|$ are Pauli operators for the qubit. Under the condition $ \nu _{1}\gg \lambda ,K,\delta$ with $\delta =\omega
_{p}-\omega _{0}-2\nu _{1}$, the resonator interacts with the
qubit at the second sideband associated with the first modulation, while the
drive works at the carrier, as shown in Fig.~1(a).

The effective dynamics can be well understood in terms of the motion of a
spin-1/2 in magnetic fields. As shown in Fig.~1(b), the transverse drive can be thought of as a
static magnetic field of strength $B_{0}=2KJ_{0}(\mu)$ along the $x$-axis, forcing the
Bloch vector of the qubit to make a Larmor precession with angular frequency $%
B_{0}$, where $J_{m}(\mu)$ denotes the $m$th Bessel function of the first kind with $\mu=\varepsilon_{1}/\nu_1$. 
The second longitudinal modulation acts as the combination of two components: $B_{1}^{\pm}$ that have the same amplitude $|B_{1}^{\pm}|=\varepsilon_{2}/4$, but rotate with opposite angular velocities $\pm\nu_{2}$ on the $yz$-plane. 
On the other hand, the quantized light field behaves like a magnetic field with the $x$- and $y$-components%
\begin{eqnarray}
	B_{2}^{x}=2\eta \left(a+a^{\dag}\right), \ \ \  \ B_{2}^{y}=2i\eta \left(a^{\dag}-a\right), 
\end{eqnarray}
where $\eta=\lambda J_{2}(\mu)/2$. 
When $B_{0}=\nu_{2}\gg B_{1},~B_{2}^{x(y)}$, in the framework coinciding with the Larmor precession, the components $B_{1}^{-}$ and $B_{2}^{y}$ can be discarded due to fast rotations [see Fig.~1(c)].
Consequently, 
the dynamics can be described by the effective quantum Rabi Hamiltonian ($\hbar=1$)
\begin{eqnarray}\label{eq3}
	H_{R}=\frac{1}{2}\Omega \sigma _{z}+\delta a^{\dagger }a+\eta \sigma _{x}\left(a+a^{\dag}\right), 
\end{eqnarray}
which is obtained by subsequently performing the transformations $\exp\left[i\int_{0}^{t}H_{0}dt\right]$ and $\exp\left(iB_0\sigma_{x}t/2 \right)$
and neglecting the fast-oscillating terms (see Supplementary Material Sec. S1~A \cite{supp}\nocite{Song_Nature_Comm2017,Ning_prl2019,song10q,duanarxiv,Pnfit2,Pnfit3,Pnfit4,cvx,nega,SchleichBook,Tavis1968,Song2019,Dicke_SPT}), where $\Omega =\varepsilon_{2}/2$ and%
\begin{eqnarray}
	H_{0}=\left(\omega _{0}+2\nu _{1}\right)a^{\dagger }a+\frac{1}{2}[\omega _{0}+\varepsilon_{1}\cos (\nu _{1}t)]\sigma_{z}.
\end{eqnarray}
We note that the synthesized qubit-resonator system corresponds to an isomorphism of the QRM \cite{Gutierrez2021}, where the effective counter-rotating-wave coupling is produced by the external drive, but not inherent in the qubit-resonator interaction as in the QRM without driving. With this realization, the system frequencies are replaced by the transverse driving detuning and the longitudinal modulation amplitude, which can be easily tuned. Consequently, the critical point of the SPT can be reached in the effective QRM without requiring the qubit-resonator coupling to be comparable to the system frequencies, thereby circumventing the restriction of the no-go theorem.



Our experimental device possesses a bus resonator and five frequency-tunable Xmon qubits, one of which is used as the test qubit for realizing the QRM. 
Before the experiment, each qubit is initialized to its ground state. The experiment starts by tuning the test qubit to the operating frequency $\omega_{0}/2\pi=5.18$~GHz, 
where a continuous microwave $K$ is applied. This transverse driving, together with the two longitudinal sine modulations, effectively realizes the Rabi Hamiltonian of Eq. (\ref{eq3}). 
The experimental details are shown in Supplementary Material Sec. S2 \cite{supp}, {including experimental setup, device parameters, and pulse sequence}. 
During the quenching process where the ratio  $\xi=2\eta/\sqrt{\Omega \delta}$ is slowly increased, all the qubits, except the test one, remain in their ground states as they are detuned from the resonator by an amount about twenty times larger than the corresponding qubit-resonator coupling strengths. 
{Such a process is realized by varying the control parameter as $\xi(t)=1.5 - \exp(-8t/t_f)$, with $t_f=2$ $\mu$s.}

To characterize the photon-number populations after a preset quench time, the microwave drive and the frequency modulations are switched off, so that the test qubit is effectively decoupled from the resonator since the detuning between the qubit and the resonator is twenty times their coupling strength without modulations.
Subsequently, an ancilla qubit is tuned on resonance with
the resonator, undergoing photon-number-dependent Rabi oscillations. 
The photonic populations of the
resonator can be inferred from the measured Rabi oscillations signals \cite{Hofheinz2009}.
Figure 2 shows the measured average photon number ($\bar{n}=\langle a^{\dag}a\rangle$) versus time. 
We compare the measured values with the theoretical predictions and show that the experimental result agrees well with the simulation.


\begin{figure*}
\centering
\includegraphics[width=18cm]{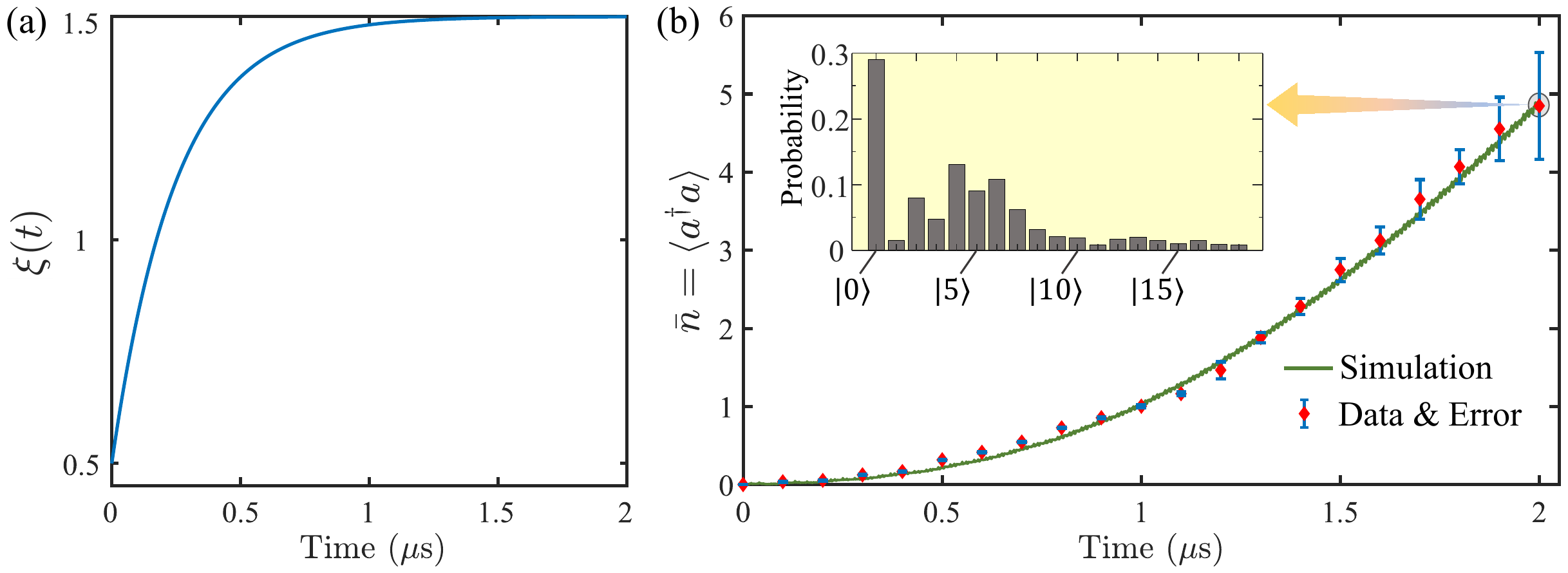}
\caption{{(a) Control parameter $\xi(t)$ versus time $t$. During the quenching process, the effective frequencies of the qubit and the resonator are respectively varied as $\Omega=2\sqrt{10}\eta/\left[1.5-\exp\left(-8t/t_{f}\right)\right]$ and $\delta\simeq\Omega/10$, with $t_{f}=2$ $\mu$s, while the effective coupling strength is fixed to $\eta/2\pi= 0.81$~MHz. Experimentally, $\Omega$ is controllable by $\varepsilon_{2}$, and $\delta$ is adjustable by the Stark shift produced by an ancilla qubit dispersively coupled to the resonator. With these settings, $\xi(t)$ depends on $t$ as $\xi(t)=1.5-\exp(-8t/t_f)$. (b)} Observed dynamical evolution of the average photon number $\bar{n}=\langle a^{\dag} a\rangle$. 
The green curve shows the result of numerical simulation based on the master equation, where the parameters of the control fields are set to: $K/(2\pi)=19.9$ MHz, $\varepsilon_1/(2\pi)=165.85$ MHz, $\nu_1/(2\pi)=200$ MHz, $\varepsilon_2=3.08 ~\Omega$, $\nu_2/(2\pi)=33.28$ MHz, and the qubit and resonator frequencies are $\omega_{0}/(2\pi)=5.18$ GHz and $\omega_{p}=\omega_{0}+2\nu_1+\delta$, respectively. The relaxation time $T_1$ (dephasing time $T_2^*$) for the qubit and the resonator are 21.5 $\mu$s and 12.9 $\mu$s (1.1 $\mu$s and 234.5 $\mu$s), respectively, each measured in independent experiments.
The inset shows the photon number distribution of the resonator measured at $t=2$ $\mu$s.}
\label{fig2}
\end{figure*}

The exotic behavior in the SP can be characterized by the
Wigner function matrix that contains full information about the joint qubit-resonator
state \cite{Wallentowitz1997}. In terms of the qubit basis $\left\{ \left\vert g\right\rangle
,\left\vert e\right\rangle \right\} $, the density operator is
expressed as
\begin{eqnarray}
	\rho=\sum_{k=g,e}\sum_{k'=g,e}\rho_{k,k'}\otimes|k\rangle\langle k'|,
\end{eqnarray}
with $\rho _{k,k'}=\left\langle k|\rho|k'\right\rangle$ matrix elements.

The information of the element $\rho_{k,k'}$ is contained in the corresponding Wigner matrix element $W_{k,k'}(\beta)$.
To measure the Wigner matrix
elements, we translate in phase space the resonator state by $\beta$. 
The matrix elements are inferred by measuring the test qubits along three mutual axes, and correlating the outcomes to the photon number distributions of the resonator measured with the ancilla qubit
(see Supplementary Material Sec. S5 \cite{supp} {for detailed characterization of the qubit-resonator state}). Figures 3(a--d) show the Wigner matrix elements reconstructed at $t=1.946~\mu$s, which reveal that the field exhibits two quasi-classical 
components with the same amplitude but opposite phases $|\pm\alpha\rangle$ and a vacuum component, featuring a first-order phase transition.
The strong quantum coherence between the ``empty" state $|0 \rangle$ and ``filled" states $|\pm\alpha \rangle$, distinguishes this SPT from the first-order phase transition previously investigated in the Dicke model with the mean field description \cite{Ferri2021,Zhu2020,Soriente2018}. 
The resulting output state can be regarded as a super-cat state, featuring being simultaneously ``empty" and ``filled", where the ``filled" state itself is a cat state composed of the two components $|\pm\alpha \rangle$ superimposed with each other.
{This super-cat state is significantly distinct from the ground state of the ideal Rabi model with an infinite frequency ratio \cite{Hwang2015}, in which the population of the vacuum component tends to 0. As shown in Fig. S7 of the Supplementary Material \cite{supp} Sec. S4, the limitation of the effective frequency ratio, non-adiabaticity, and deviation from model Hamiltonian contribute vacuum populations of 0.04, 0.03, 0.06, respectively. The dissipation is responsible for the rest of vacuum population ($\sim$0.16), and turns the qubit-resonator system into a partially mixed state, with a purity of ${\rm Tr}(\rho^2 )=0.46$. With the increase of the distance from the critical point, the size of the cat state would be improved at the price of degradation of the state purity, as a consequence of its increasing sensitivity to dissipation and the longer time needed for the quenching process.}

\begin{figure*}
	\centering
	\includegraphics[width=18cm]{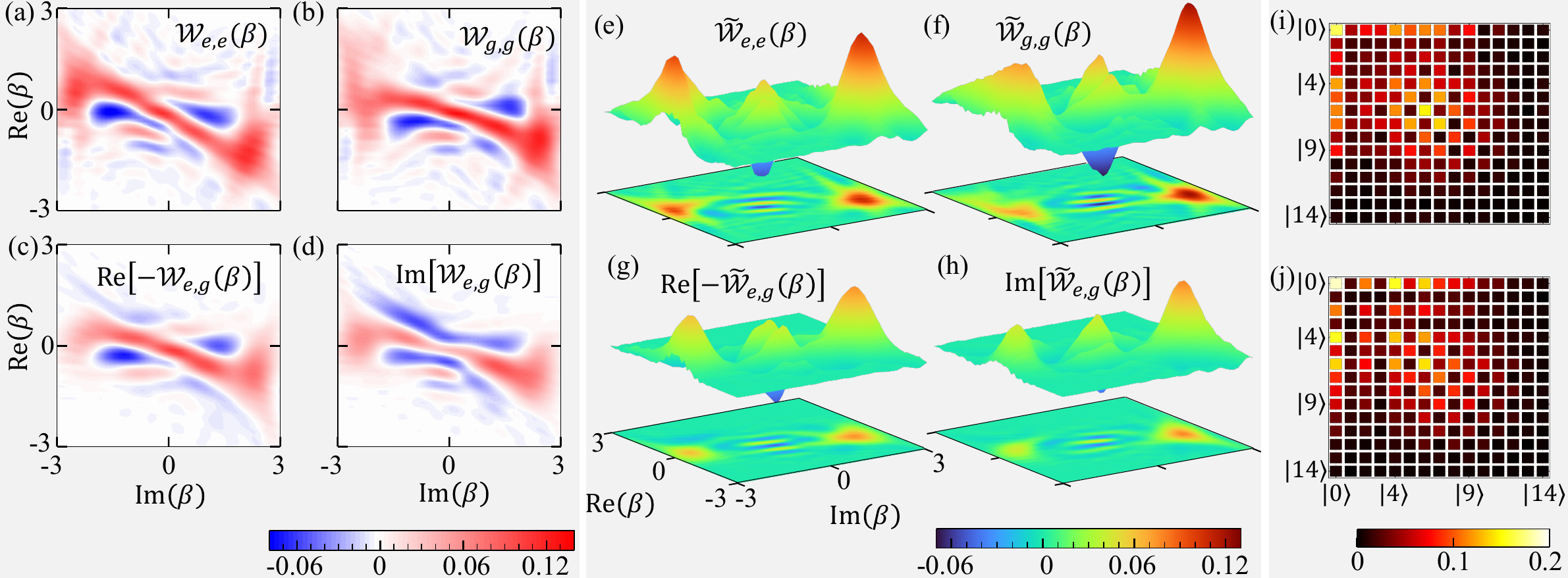}
	\caption{Wigner matrix tomography. (a--d) Measured matrix elements for the super-cat state. The value of the matrix elements at each point is reconstructed from the Wigner diagonal elements measured along the three mutually orthogonal axes of the Bloch sphere of the qubit. Each of these diagonal elements is obtained from the measured photon number distribution of the field displaced by $\beta$ in phase space, correlated with detection of the test qubit along the corresponding axis. All the data are measured at $t=1.946~\mu$s.  (e--h) Matrix elements for the superradiant-phase cat state. {These Wigner matrix elements are extracted by reconstructing the qubit-resonator density matrix in the Fock basis, discarding the elements associated with $|0\rangle$ to obtain system density matrix associated with the superradiant phase, from which the corresponding Wigner matrix is obtained.} The inferred vacuum population in the superradiant phase is $\sim$0.1\%, which implies that the measured vacuum population of about 30\% mainly arises from the normal phase  that coexists with the superradiant phase. (i,~j) Normalized density matrices of the resonator associated with the qubit states $|e\rangle$ and $|g\rangle$. For clarity, we here only display the magnitudes of the elements.
	}
	\label{fig3}
\end{figure*}

To further confirm this exotic phenomenon, we separate the measured coherent 
fields with amplitude $|\alpha|$ from the vacuum field, enabled by their large distance in phase space. The extracted Wigner matrix elements for the SP are displayed in Figs.~3(e--h). 
As expected, each of these elements exhibits two peaks,
between which there exists an oscillating pattern featuring the alternating
appearance of positive and negative values, as a consequence of quantum
interference between $\left\vert \pm \alpha \right\rangle$ \cite{Delglise2008}. The distortions
of the Gaussian peaks are mainly due to the neglected high-order nonlinear
processes and the limited frequency ratio $\Omega/\delta\simeq 10$. These results unambiguously demonstrate that the photons produced
in the SP spontaneously form a cat-like state. 
For each of the two quasiclassical components forming the cat state, the parity symmetry is broken. The extent of the symmetry breaking is quantified by the field coherence, $\langle a \rangle$, which is equal to the amplitude of the quasiclassical coherent state and can be used as an order parameter to characterize the QPT \cite{Hwang2015,Hwang2016}.

After the quenching dynamics, this coherence has a magnitude of 2.62, which indicates the occurrence of a phase transition during the quenching process \cite{Sachdev2011}. The resulting cat state formed by two coherent states $\vert \pm \alpha \rangle$ has a size of $d^2=4|\alpha|^2=27.46$ \cite{Delglise2008}.
The inferred diagonal element $\tilde{\mathcal{W}}_{e,e}$ ($\tilde{\mathcal{W}}_{g,g}$) has a minimum value of $-$0.060 ($-$0.068) at $\beta=-0.48-0.36i$ ($\beta=-0.48i$). These negative phase-space quasi-probability densities show the nonclassicality of the emergent photonic field. 
The photon-matter entanglement can be quantified by using the negativity obtained from the partially transposed density matrix \cite{Horodecki2009}. {The negativity for the SP, inferred from the qubit-resonator density matrix associated with the SP, is 0.12, which confirms the existence of strong light-matter entanglement, making the present SP distinct from those realized in previous experiments \cite{Zhang2017,Baumann2010,Baumann2011,Brennecke2013,Klinder2015,Leonard2017,Ferri2021,Zhang2021}, where no entanglement was observed due to the strong decoherence. Because of the decoherence effects, this negativity is smaller than that for the ideal superradiant ground state, which has a value of 0.44.}

The quantum coherence between the coexisting
phases can be further confirmed by the off-diagonal elements between $|0\rangle$ and $|n\rangle$ $(n\neq0)$ in the Fock basis associated with ${\cal W}_{e,e}$ and ${\cal W}_{g,g}$, displayed in 
Figs.~3(i) and (j). The coherences between the empty state and filled state, defined as {$\mathcal{C}_{k,k}=\sum_{n\neq0}|\langle 0|\rho_{k,k}|n\rangle|/{\rm Tr}(\rho_{k,k})$}, 
are 1.018 and 1.020 for the renormalized resonator density matrices correlated with $|e\rangle$ and $|g\rangle$, respectively. 
Each of these coherences is much larger than that of the coherent state $|\alpha\rangle$, {$\mathcal{C}_{\alpha}=\sum_{n\neq 0} |\langle 0|\alpha\rangle\langle\alpha|n\rangle|=0.1147$}, 
{which verifies that these coherences are mainly due to the quantum superposition between the SP and the normal phase}. 
{The negativity of the realized first-order phase transition is 0.25, which quantifies the qubit-resonator of the output density matrix, reconstructed without removing the elements associated with the vacuum state component $|0\rangle$. This negativity is significantly higher than that for the SP, mainly due to the fact that the vacuum state is not subjected to decoherence.}

In conclusion, we have theoretically proposed and experimentally demonstrated a method for exploring the SPT of a microwave photonic field stored in a resonator coupled to a superconducting artificial atom. 
The reconstructed resonator-qubit Wigner matrix reveals quantum interference effects between the vacuum and the SP cat states, and between the two SP states, as well as light-matter entanglement. 
It is expected that the emergent cat state loses its coherence at an increasing rate with the increase of the photon number, as a consequence of the information acquisition about the phase or amplitude of the field by the environment.
Progressively increasing the quench parameter would make it possible to experimentally explore the intimate relation between the symmetry breaking process and decoherence, which plays a central role in the quantum-to-classical transition. 
In addition to fundamental interest, our system may find applications in quantum technology.


\begin{acknowledgments}
We acknowledge helpful discussions with Anton Frisk Kockum. This work was supported by the National
Natural Science Foundation of China (Grants No. 12274080, No. 11875108, No. 11934018, No. 11904393,
No. 92065114, and No. T2121001), Innovation Program for Quantum Science and Technology under Grant No. 2021ZD0300200, the Strategic Priority
Research Program of Chinese Academy of Sciences
(Grant No. XDB28000000), 
the Key-Area Research and Development Program of Guangdong Province, China (Grant No. 2020B0303030001), 
the Natural Science Funds for Distinguished Young Scholar of Fujian Province under Grant No. 2020J06011, Projects from Fuzhou University under Grants No. JG202001-2 and No. 049050011050,
and Beijing Natural Science Foundation (Grant No. Z200009).
A.M. is supported by the Polish National Science Centre (NCN) under the Maestro Grant No. DEC-2019/34/A/ST2/00081.
F.N. is supported in part by:
Nippon Telegraph and Telephone Corporation (NTT) Research,
the Japan Science and Technology Agency (JST) [via
the Quantum Leap Flagship Program (Q-LEAP),
and the Moonshot R\&D Grant No.~JPMJMS2061],
the Asian Office of Aerospace Research and Development (AOARD) (via Grant No.~FA2386-20-1-4069), and
the Foundational Questions Institute Fund (FQXi) via Grant No.~FQXi-IAF19-06.
\end{acknowledgments}

\bibliography{references}

\end{document}


\title{Supplemental Material for\\ ``Observation of a superradiant phase transition with emergent cat states''}

	\author{Ri-Hua Zheng}
	\thanks{These authors contribute equally to this work.}
	\author{Wen Ning}
	\thanks{These authors contribute equally to this work.}
	\affiliation{Fujian Key Laboratory of Quantum Information and Quantum
		Optics, College of Physics and Information Engineering, Fuzhou University,
		Fuzhou, Fujian 350108, China}
	\author{Ye-Hong Chen}
	\thanks{These authors contribute equally to this work.}
	\affiliation{Fujian Key Laboratory of Quantum Information and Quantum
		Optics, College of Physics and Information Engineering, Fuzhou University,
		Fuzhou, Fujian 350108, China}
	\affiliation{Theoretical Quantum Physics Laboratory, RIKEN Cluster for Pioneering 
		Research, Wako-shi, Saitama 351-0198, Japan}
	\affiliation{Quantum Information Physics Theory Research Team, RIKEN Center for Quantum Computing (RQC), Wako-shi, Saitama 351-0198, Japan}
	
	\author{Jia-Hao L\"{u}} 
	\affiliation{Fujian Key Laboratory of Quantum Information and Quantum
		Optics, College of Physics and Information Engineering, Fuzhou University,
		Fuzhou, Fujian 350108, China}
	
	\author{Li-Tuo Shen} 
	\affiliation{Fujian Key Laboratory of Quantum Information and Quantum
		Optics, College of Physics and Information Engineering, Fuzhou University,
		Fuzhou, Fujian 350108, China}
	
	\author{Kai Xu}
	\affiliation{Institute of Physics and Beijing National Laboratory for Condensed
		Matter Physics, Chinese Academy of Sciences, Beijing 100190, China}
	\affiliation{CAS Center for Excellence in Topological Quantum Computation,
		University of Chinese Academy of Sciences, Beijing 100190, China}
	
	\author{Yu-Ran Zhang}
	\affiliation{School of Physics and Optoelectronics, South China University of Technology, Guangzhou 510640, China}
	\affiliation{Theoretical Quantum Physics Laboratory, RIKEN Cluster for Pioneering Research, Wako-shi, Saitama 351-0198, Japan}
	\affiliation{Quantum Information Physics Theory Research Team, RIKEN Center for Quantum Computing (RQC), Wako-shi, Saitama 351-0198, Japan}

	\author{Da Xu}
	\affiliation{Interdisciplinary Center of Quantum Information, State Key Laboratory of Modern Optical Instrumentation
		and Zhejiang Province Key Laboratory of Quantum Technology and Device, School of Physics, Zhejiang University,
		Hangzhou 310027, China}
	
	\author{Hekang Li}
	\affiliation{CAS Center for Excellence in Topological Quantum Computation,
		University of Chinese Academy of Sciences, Beijing 100190, China}
	

	\author{Yan Xia}
	\affiliation{Fujian Key Laboratory of Quantum Information and Quantum
		Optics, College of Physics and Information Engineering, Fuzhou University,
		Fuzhou, Fujian 350108, China}
	
	\author{Fan Wu}
	\affiliation{Fujian Key Laboratory of Quantum Information and Quantum
		Optics, College of Physics and Information Engineering, Fuzhou University,
		Fuzhou, Fujian 350108, China}
	
	\author{Zhen-Biao Yang}\email{zbyang@fzu.edu.cn}
	\affiliation{Fujian Key Laboratory of Quantum Information and Quantum
		Optics, College of Physics and Information Engineering, Fuzhou University,
		Fuzhou, Fujian 350108, China}
	
	\author{Adam Miranowicz}
	\affiliation{Theoretical Quantum Physics Laboratory, RIKEN Cluster for Pioneering Research, Wako-shi, Saitama 351-0198, Japan}
	\affiliation{Institute of Spintronics and Quantum Information, Faculty of Physics, Adam Mickiewicz University, 61-614 Pozna\'{n}, Poland}
	
	\author{Neill Lambert}
	\affiliation{Theoretical Quantum Physics Laboratory, RIKEN Cluster for Pioneering 
		Research, Wako-shi, Saitama 351-0198, Japan}
	
	\author{Dongning Zheng}
	\affiliation{Institute of Physics and Beijing National Laboratory for Condensed
		Matter Physics, Chinese Academy of Sciences, Beijing 100190, China}
	\affiliation{CAS Center for Excellence in Topological Quantum Computation,
		University of Chinese Academy of Sciences, Beijing 100190, China}
	
	\author{Heng Fan}
	\affiliation{Institute of Physics and Beijing National Laboratory for Condensed
		Matter Physics, Chinese Academy of Sciences, Beijing 100190, China}
	\affiliation{CAS Center for Excellence in Topological Quantum Computation,
		University of Chinese Academy of Sciences, Beijing 100190, China}
	
	\author{Franco 	Nori}\email{fnori@riken.jp}
	\affiliation{Theoretical Quantum Physics Laboratory, RIKEN Cluster for Pioneering 
		Research, Wako-shi, Saitama 351-0198, Japan}
	\affiliation{Quantum Information Physics Theory Research Team, RIKEN Center for Quantum Computing (RQC), Wako-shi, Saitama 351-0198, Japan}
	\affiliation{Department of Physics, University of Michigan, Ann Arbor, Michigan
		48109-1040, USA}

	\author{ Shi-Biao Zheng}\email{t96034@fzu.edu.cn}
	\affiliation{Fujian Key Laboratory of Quantum Information and Quantum
		Optics, College of Physics and Information Engineering, Fuzhou University,
		Fuzhou, Fujian 350108, China}

\maketitle

\tableofcontents

\newpage


\section{Effective quantum Rabi model}\label{ss1}

{	\renewcommand\arraystretch{1.2}
\begin{table}[b] 
	\centering
	\begin{tabular}{|p{3cm}<{\centering}|p{7cm}<{\centering}|}
		\hline
		ADC & {Analog-to-Digital Converter}\\
		\hline
		DAC & {Digital-to-Analog Converter} \\
		\hline
		DC & {Direct Current} \\
		\hline
		HEMT & High { Electron Mobility Transistor} \\
		\hline
		{IQ}  & {In-phase and Quadrature} \\
		\hline
		JPA & Josephson {Parametric Amplifier} \\
		\hline
		MC & Mixing {Chamber}\\
		\hline
		MS & Microwave {Source} \\      
		\hline
		NP&Normal {Phase}  \\   
		\hline
		RC & Resistance Capacitance \\
		\hline
		RI & {Read-In} \\
		\hline
		RO & {Read-Out}\\
		\hline
		SPT & Superradiant {Phase Transition} \\
		\hline
		SP& Superradiant {Phase} \\
		\hline
	\end{tabular}
	\caption{\label{abb} Abbreviations {used} in this supplemental material.}
\end{table}
}

\subsection{Derivation of the effective quantum Rabi Hamiltonian}
When the drive is tuned to the carrier, the
dynamics is described by the Hamiltonian ($\hbar=1$ hereafter)

\begin{eqnarray}
H=H_{0}+H_{I}, 
\end{eqnarray}%
where%

\begin{eqnarray}
H_{0} &=&\left(\omega _{0}+2\nu _{1}\right)a^{\dagger }a+\frac{1}{2}[\omega _{0}+\varepsilon_{1}\cos (\nu _{1}t)]\sigma_{z}, \cr\cr
H_{I} &=&\delta a^{\dagger }a+\frac{1}{2}\varepsilon _{2}\cos \left(\nu
_{2}t \right)\sigma_{z} +\left (\lambda
a^{\dagger }\sigma_{-}+K
e^{i\omega_0 t}\sigma_{-} +{\rm{h.c.}}\right).
\end{eqnarray}%
Here, $\sigma_{z}=|e\rangle\langle e|-|g\rangle\langle g|$ and $\sigma_{-}=|g\rangle\langle e|$ are Pauli operators for the qubit.
Performing the transformation 

\begin{eqnarray}
U_{0}=\exp{\left[i\int_{0}^{t}H_{0} \ dt\right]},
\end{eqnarray} 
we obtain the
system Hamiltonian in the rotating frame%

\begin{eqnarray}
H_{I}^{\prime }=\delta a^{\dagger }a+\frac{1}{2}\varepsilon _{2}\cos (\nu
_{2}t)\sigma_{z}+\left\{\exp \left[ -i\mu \sin (\nu _{1}t)\right] \left[\lambda \exp
\left( 2i\nu _{1}t\right) a^{\dagger }+K \right]\sigma_{-} +{\rm{h.c.}}\right\} ,
\end{eqnarray}%
where $\mu =\varepsilon _{1}/\nu _{1}$. Using the Jacobi-Anger
expansion

\begin{eqnarray}
\exp \left[ i\mu \sin \left( \nu _{1}t\right) \right] =\stackrel{\infty }{%
\mathrel{\mathop{\sum }\limits_{m=-\infty }}%
}J_{m}(\mu )\exp \left( im\nu _{1}t\right) , 
\end{eqnarray}%
with $J_{m}(\mu )$ being the $m$th Bessel function of the first kind, we
obtain%

\begin{eqnarray}\label{eqS6}
H_{I}^{\prime } &=&\delta a^{\dagger }a+\frac{1}{2}\varepsilon _{2}\cos (\nu
_{2}t)\sigma_{z}+\left(\stackrel{\infty }{%
	\mathrel{\mathop{\sum }\limits_{m=-\infty }}%
}J_{m}(\mu )\left\{\lambda \exp [-i(m-2)\nu _{1}t]a^{\dagger }+K \exp \left(
-im\nu _{1}t\right) \right\}\sigma_{-}
+{\rm{h.c.}}\right).
\end{eqnarray}
{Assuming} that $\left\{\left\vert \lambda J_{2}(\mu )\right\vert,~\left\vert
K J_{0}(\mu )\right\vert,~\delta,~\varepsilon_2\right\} \ll \nu
_{1}$, we can discard the {fast-oscillating} terms, thus $%
H_{I}^{\prime }$ reduces to

\begin{eqnarray}
H_{I}^{\prime } &=&\left[J_{2}(\mu )\lambda a^{\dagger }\sigma_{-} +K J_{0}(\mu
)\sigma_{-} +{\rm{h.c.}} \right]+ \delta a^{\dagger }a+\frac{1}{2}\varepsilon _{2}\cos (\nu
_{2}t)\sigma_{z} \cr\cr
&=&\frac{1}{2}B_{0}\sigma _{x}+\frac{1}{2}\varepsilon
_{2}\cos (\nu _{2}t)\sigma _{z}+\delta a^{\dagger }a +\eta \left(X\sigma _{x}-Y\sigma _{y}\right),
\end{eqnarray}
where $B_{0}=2KJ_{0}(\mu )$, $\eta =\lambda J_{2}(\mu )/2$, $X=a^{\dagger
}+a $, $Y=i(a^{\dagger }-a)$, $\sigma_{x}=\sigma_{-}+\sigma_{-}^{\dag}$, and $\sigma_{y}=i\sigma_{-}-i\sigma_{-}^{\dag}$. Under the transformation $\exp (iB_0\sigma _{x}t/2)$, the system
Hamiltonian in the moving frame becomes 

\begin{eqnarray}\label{eqS8}
H_{I}^{\prime \prime } &=&\eta \left\{X \sigma _{x}-Y[\cos (B_0t)\sigma
_{y}-\sin (B_0t)\sigma _{z}]\right\} \cr\cr
&&+\;\delta a^{\dagger }a+\frac{1}{2}\varepsilon _{2}\cos (\nu
_{2}t)\left[\cos (B_0t)\sigma _{z}+\sin (B_0t)\sigma _{y}\right].
\end{eqnarray}%
In the limit of $B_0\gg \eta ,\varepsilon _{2}/2$ and $\nu _{2}=B_{0}$, the {fast-oscillating} terms can be neglected, and $H_{I}^{\prime \prime }$ reduces to
the Rabi Hamiltonian%

\begin{eqnarray}\label{eqS9}
H_{R}=\frac{1}{2}\Omega \sigma _{z}+\delta a^{\dagger }a+\eta \sigma
_{x}(a^{\dagger }+a), 
\end{eqnarray}%
where $\Omega=\varepsilon _{2}/2$ is the effective qubit frequency.

Some fast-oscillating terms in Eq.~(\ref{eqS6}) cannot be  ignored, and induce
additional Stark shifts to the effective qubit- and cavity-frequencies. 
Detailed discussions about these Stark shifts are shown in Sec.~\ref{Stark}.

{\subsection{Superradiant phase transition under ideal conditions}

In the limit of $\Omega/\delta\rightarrow \infty$, we can diagonalize the Rabi Hamiltonian $H_{R}$ using a Schrieffer-Wolff {transformation} \cite{2015PRL}.
After applying a unitary operator

\begin{align}
U_{\rm{SW}}=\exp{\left[i\frac{\eta}{\Omega}\left(a^{\dag}+a\right)\sigma_{y}\right]},
\end{align}
and keeping the terms {up to}
second order {in} the qubit-resonator coupling strengths, Eq.~(\ref{eqS9}) becomes 

\begin{align}\label{eqS11}
H_{\rm np}=\delta a^{\dag}a+\frac{\Omega}{2}\sigma_{z}+\frac{\delta \xi^{2}}{4}\left(a+a^{\dag}\right)^{2}\sigma_{z},
\end{align}
which provides a faithful description of the system ground state
in the normal phase (NP) of the model, where $\xi=2\eta/\sqrt{\Omega\delta}$ is a normalized coupling parameter.
Equation (\ref{eqS11}) shows that the ground-qubit-state subspace $\left\{|n\rangle|g\rangle\right\}$ is decoupled from the excited-qubit-state subspace $\left\{|n\rangle|e\rangle\right\}$. Therefore,
upon a projection $\langle g|H_{\rm np}|g\rangle$ for $\xi\leq 1$, one can solve the ground eigenstate and eigenvalue of $H_{R}$
as 

\begin{align}
|\psi_{\rm np}\rangle=S(r_{\rm np})|0\rangle|g\rangle, \ \ \ \ \ \mathcal{E}_{\rm np}=\delta\sqrt{1-\xi^{2}},
\end{align}
where $S(r_{\rm np})=\exp{\left[\left(r_{\rm np}a^{\dag 2}-r_{\rm np}^{*}a^{2}\right)/2\right]}$ is the squeezing operator with a real squeezing parameter $r_{\rm np}=-\frac{1}{4}\ln\left(1-\xi^{2}\right)$.
Thus, the excitation energy $\mathcal{E}_{\rm np}$ is a positive real number for
$\xi<1$ and vanishes at $\xi=1$, i.e., in the NP.

For $\xi>1$, the number
of photons occupied in the cavity field becomes proportional
to $\Omega/\delta$ and acquires
macroscopic occupations, i.e., in the superradiance phase (SP) \cite{2015PRL}.
To capture
the physics of the SP,
we displace the bosonic
mode in the Rabi Hamiltonian $H_{R}$. For a displacement parameter $\alpha=\sqrt{\left[\Omega/\left(4\xi^{2}\delta\right)\right]\left(\xi^{4}-1\right)}$, we obtain 

\begin{align}
H'_{R}(\pm\alpha)=&D^{\dag}(\pm\alpha)H_{R}D({\pm\alpha})\cr
              =&\delta a^{\dag}a+\frac{\tilde{\Omega}}{2}\tilde{\sigma}_{z}^{\pm}+\tilde{\eta}(a+a^{\dag})\tilde{\sigma}_{x}^{\pm}+\delta\alpha^{2},
\end{align}
where $\tilde{\Omega}=\xi^{2}\Omega$, $\tilde{\eta}=\sqrt{\delta\Omega}/(2\xi)$, and 

\begin{align}
D(\alpha)=\exp\left(\alpha a^{\dag}-\alpha^{*}a\right), \ \ \ \ \ \ \ \  \tilde{\sigma}_{z}^{\pm}=|\uparrow^{\pm}\rangle\langle\uparrow^{\pm}|-|\downarrow^{\pm}\rangle\langle\downarrow^{\pm}|, \ \ \ \ \ \ \ \ \tilde{\sigma}_{x}^{\pm}=|\uparrow^{\pm}\rangle\langle\downarrow^{\pm}|+|\downarrow^{\pm}\rangle\langle \uparrow^{\pm}|.
\end{align}
The states 

\begin{align}
|\uparrow^{\pm}\rangle=\cos(\theta)|e\rangle\pm\sin(\theta)|g\rangle,	\ \ \ \ \ \ \  |\downarrow^{\pm}\rangle=\mp\sin(\theta)|e\rangle+\cos(\theta)|g\rangle,
\end{align}
are the eigenstates of the terms {$(\frac{\Omega}{2}\sigma_{z}\pm2\alpha\eta\sigma_{x})$} that construct a new qubit eigenstate subspace, where $\theta$ obeys

\begin{align}
\tan{\theta}=\sqrt{\frac{\xi^{2}-1}{\xi^{2}+1}}.
\end{align}
Then, employing the same procedure used to derive $H_{\rm np}$, we obtain

\begin{align}\label{eqS16}
H_{\rm sp}=\delta a^{\dag}a+\frac{\delta}{4\xi^{4}}\left(a^{\dag }+a\right)^{2}\tilde{\sigma}^{\pm}_{z}+\frac{\Omega}{4}\left(\xi^{2}+\xi^{-2}\right)\tilde{\sigma}^{\pm}_{z}.
\end{align}
Applying the projection $\langle \downarrow^{\pm}|H_{\rm sp}|\downarrow ^{\pm}\rangle$, Eq.~(\ref{eqS16}) becomes

\begin{align}
H_{\rm sp}=\delta a^{\dag}a-\frac{\delta}{4\xi^{4}}\left(a^{\dag }+a\right)^{2}-\frac{\Omega}{4}\left(\xi^{2}+\xi^{-2}\right),
\end{align}
whose excitation energy is found to be $\mathcal{E}_{\rm sp}=\delta\sqrt{1-\xi^{-4}}$.
The ground eigenstates of the quantum Rabi Hamiltonian $H_{R}$ for $\xi>1$ are

\begin{align}
|\psi_{\rm sp}\rangle =D(\pm\alpha)S(r_{\rm sp})|0\rangle|\downarrow^{\pm}\rangle,
\end{align}
which are degenerate,
where $r_{\rm sp}=-\frac{1}{4}\ln\left(1-\xi^{-4}\right)$.

Therefore, $H_{\rm np}$ and $H_{\rm sp}$ are the exact low-energy
effective {Hamiltonians} for the NP $(\xi<1)$ and
SP $(\xi>1)$, respectively.

}

\section{Experimental setup, device parameters, and pulse sequence}

\begin{figure}[htb]
\centering
\includegraphics[width=15cm]{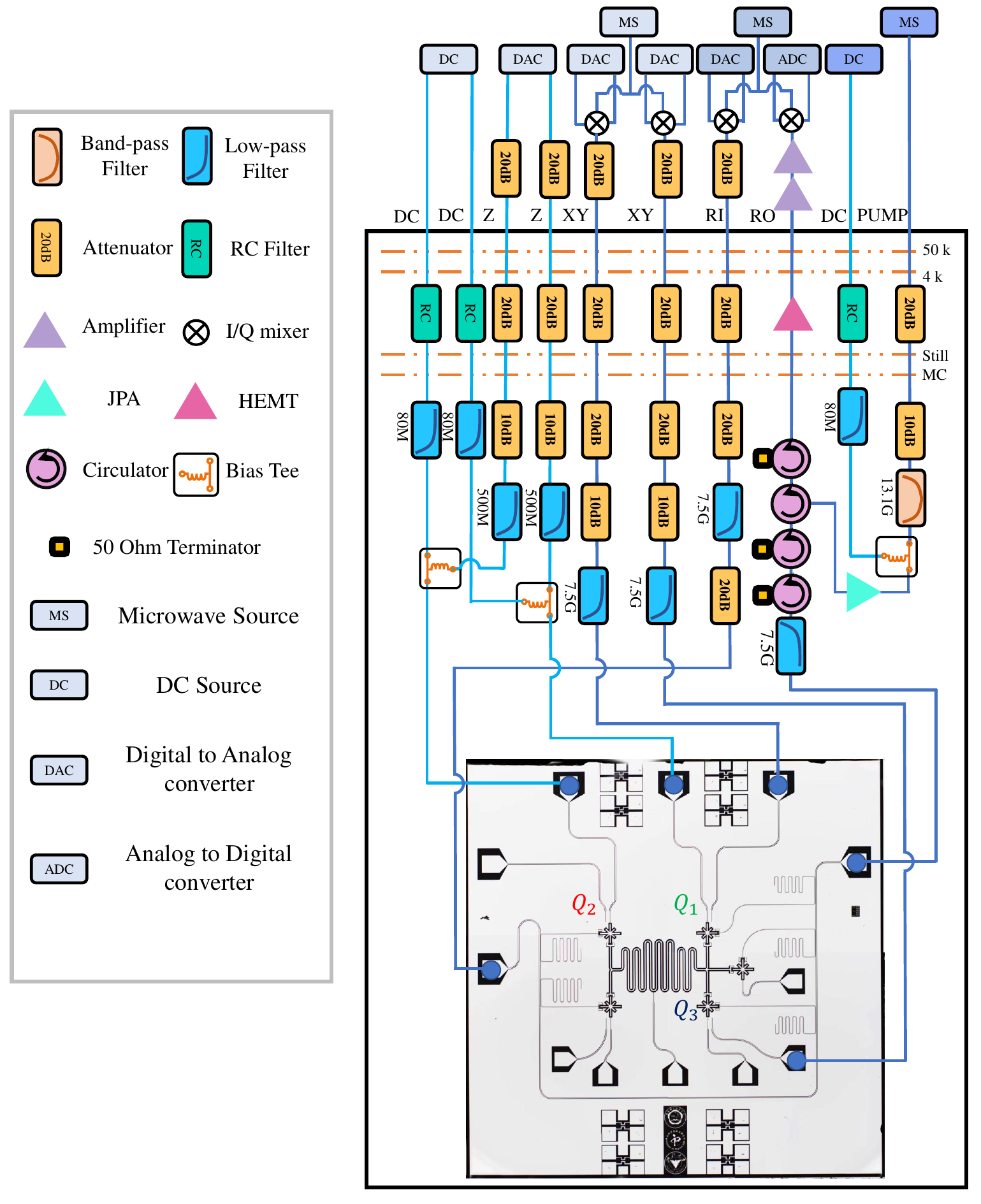}
\caption{Schematic diagram of the experimental setup. Note there is a low-pass filter (7.5G) inverted with the others because it connects to the read-out line.}
\label{wiring}
\end{figure}

The whole electronics and wiring for our superconducting circuit control are outlined in Fig. \ref{wiring} \cite{Song_Nature_Comm2017, Ning_prl2019}. 
The superconducting circuit sample, used in our experiment, possesses a bus resonator and five Xmon qubits. Each qubit has its own {read-out} resonator for reading out its states.
The bus resonator has a fixed bare frequency $\omega_p/2\pi = 5.581$ GHz and a photonic decay time $T_{1,p} =12.9$ $\mu$s. 
Every qubit has two control lines: an XY-control for flipping its states and a Z-control for modulating its frequency, allowing the qubit to flexibly couple with the bus resonator. 
This, together with the relatively long lifetime of the resonator photons, guarantees the slow quenching
manipulation of the qubit-resonator ground state to induce the appearance of the superradiant phase transition (SPT) accompanied 
by the sudden birth of a photonic mesoscopic superposition of a considerable size. 
For clarity, {Table~\ref{abb} lists the abbreviations used in this supplemental material}.

The XY-controls on the qubits are implemented through the mixing of 
the low-frequency signals yielded by the {IQ} channels of two digital-to-analog converters (DACs) and a microwave source (MS). The carrier frequency of the MS is about $5.5$ GHz. The Z-controls on the qubits are implemented by two signals: one is produced by the direct-current (DC) biasing line from a 
low-frequency DC source; the other {comes} directly from the Z-control of a DAC. 
The qubit {read-out} is realized through mixing the signals of the {IQ} channels of two DACs and an
MS with a frequency $\sim6.69$ GHz, {which output a {read-out} pulse with multiple tones targeting 
all resonators for qubit {read-out}.} 
The output from the circuit  is amplified sequentially by an impedance-transformed Josephson parametric amplifier (JPA), high electron mobility transistor (HEMT), and room temperature amplifiers.
Then, it is captured and demodulated by analog-to-digital converters (ADCs).
Both DACs and ADCs are supported by a field-programmable gate array which reacts at a nanosecond-level speed. 
The JPA is pumped by an MS with a frequency $\sim13.5$ GHz 
and modulated by a DC bias.
Moreover, some circulators, attenuators, and filters are added to the signal lines in 
specific temperature regions to reduce the noises that influence the performance of the device. 

\begin{figure}[htp]
\centering
\includegraphics[width=14cm]{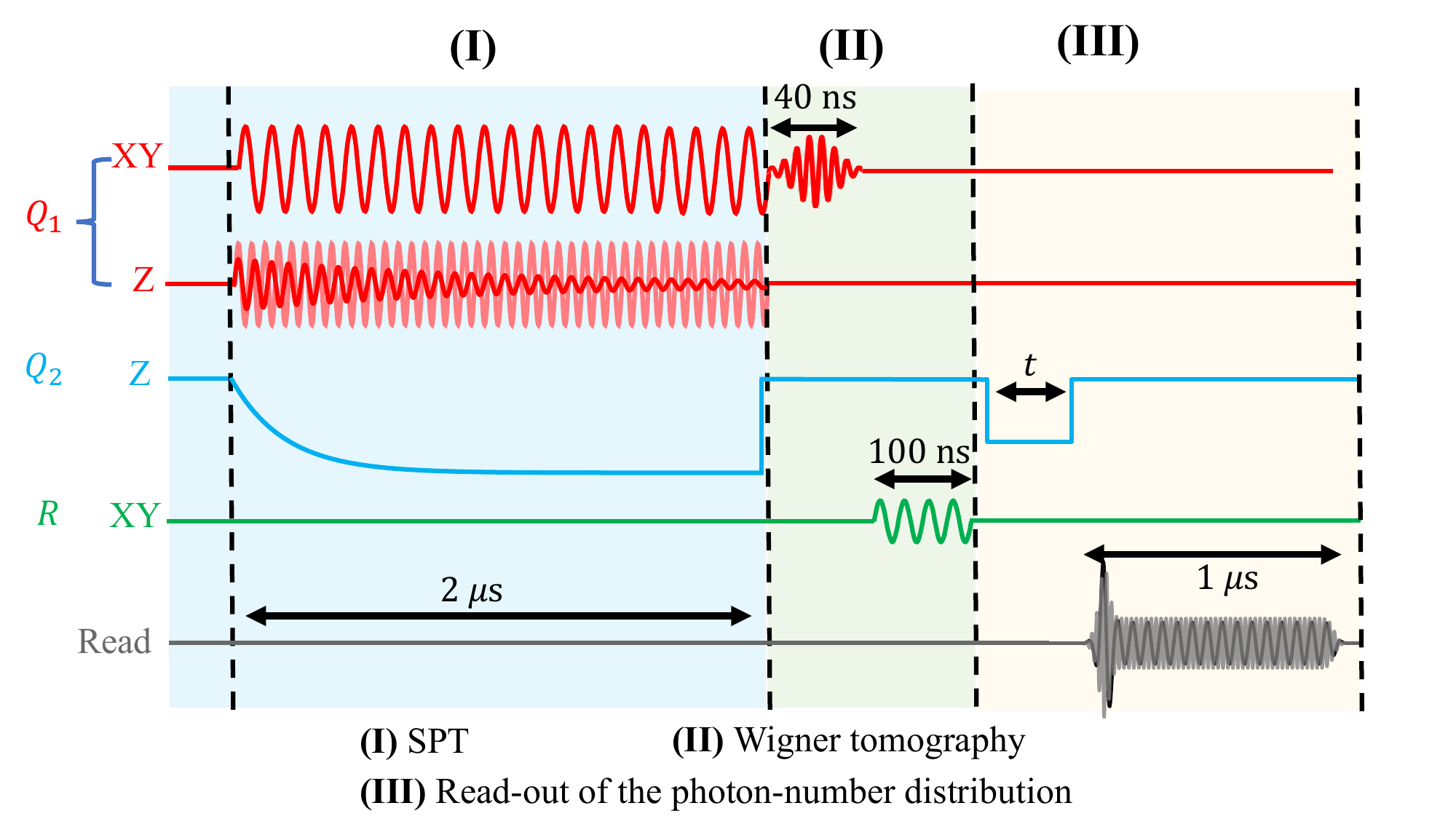}
\caption{Sketch of the pulse sequences, which consists of three steps: (I) superradiant phase transition (SPT); (II) Wigner tomography; {and} (III) {read-out} of the photon-number distribution. In (I), we apply a continuous drive pulse (amplitude $K/2\pi=19.91$ MHz and frequency $\omega_0/2\pi=5.18$  GHz) to the XY-line and two tunable pulses [$\varepsilon_1 \cos (\mu_1 t)$ and $\varepsilon_2 \cos (\mu_2 t)$, with $\{\varepsilon_1,~\nu_1,~\nu_2\}/2\pi=\{165.85,~200,~33.28\}$ MHz] to the Z-line of the test qubit $Q_1$, where the amplitude of the second modulation pulse, $\varepsilon_2$, slowly decreases with time, corresponding to the decrease of $\Omega(t)$. Moreover, the frequency $f(t)$ of the ancilla qubit $Q_2$ decreases with the amplitude of the Z-line, inducing a decrease in the $\delta(t)$ by Stark shifts. In (II) a {single-qubit} rotation operation [the identity operation, $R_x(\pi/2)$, or $R_y(\pi/2)$] is applied to the test qubit, and a displacement operation $D(-\beta)$ is then applied to the resonator (actually by tuning the XY-line of $Q_3$ with cross-talk interactions, not shown). In (III) the ancilla qubit $Q_{2}$ is resonantly coupled to the resonator for a given time $\tau$, and then biased to its idle frequency ($f_{\rm idle}/2\pi = 5.93$ GHz) for state {read-out}. }
\label{sequence}
\end{figure}

\begin{table} 
\centering 
\begin{tabular}{ccccccccc}
	\hline
	\hline
	& \ \ \ \multirow{2}{*}{$ \omega_{10}/2\pi$ (GHz)}  \ \ \ 
	&\ \ \ \multirow{2}{*}{$T_{1}$ ($\mu$s)} \ \ \  
	& \ \ \ \multirow{2}{*}{$T_{2}^*$ ($\mu$s)} \ \ \  
	&\ \ \ \multirow{2}{*}{$T_{2}^{\textrm{SE}}$ ($\mu$s)}\ \ \ 
	& \ \ \ \multirow{2}{*}{$\lambda/2\pi$ (MHz)}  \ \  \ 
	&\ \ \ \multirow{2}{*}{$\gamma/2\pi$ (MHz)} \ \ \  
	& \ \ \ \multirow{2}{*}{$F_{g}$} \ \ \ &\multirow{2}{*}{$F_{e}$}\\
	~\\
	\hline
	$Q_1$&5.180&21.5&1.1&6.0&19.91&250&0.983&0.937\\
	$Q_2$&5.930&17.2&1.5&14.3&20.92&238&0.990&0.920\\
	\hline
	$R$&5.581&12.9&234.5&-&-&-&-&-\\
	\hline
	\hline
\end{tabular}
\caption{\label{parameter} \textbf{Qubit and resonator characteristics.} {The symbols} $Q_1$, $Q_2$, and $R$ correspond to {the} test qubit, the ancilla qubit, and the resonator, respectively. The idle frequencies of $Q_j$ ($j=1,2$) and $R$ are generally marked by $\omega_{10}/2\pi$, where single-qubit rotation pulses and {measurements} are applied. 
For the decoherence performance, $T_{1}$ and $T_{2}^*$ are the energy relaxation time and the Ramsey dephasing time (Gaussian decay), respectively, of $Q_j$ and $R$,  measured at the idle frequency. Additionally, $T_{2}^{\textrm{SE}}$ is the dephasing time with spin echo (Gaussian decay). The coupling strength $\lambda$ between $Q_j$ and the bus resonator $R$ is estimated via vacuum Rabi oscillations. The anharmonicity of {the} qubit {is} $\gamma$. The probability of detecting {the} qubit in $\vert g\rangle$ ($\vert e\rangle$) when it is prepared in $\vert g\rangle$ ($\vert e\rangle$) state is indicated by $F_{g}$ ($F_{e}$).}
\label{par}
\end{table}

Three of the five qubits are used in our experiment.
One is used as the test qubit $Q_1$ for realizing the effective quantum Rabi model. 
The second one acts as an ancilla qubit $Q_2$ to determine the photon-number distribution for {analyzing the Wigner function distribution whose negativity reveals and quantifies the nonclassicality of the light field.}
The XY-line of the third qubit $Q_3$ is used 
to control the bus resonator (by cross-talk interactions) for performing a displacement operation on its states in phase space.  

The performance characterization of the qubits and the resonator are listed in Table \ref{parameter}. For technical details about the superconducting qubits, e.g., see Ref. \cite{song10q}, which reports similar control methods to our experiment.

The pulse sequences are shown in Fig. \ref{sequence}, including three steps: (I) SPT; (II) Wigner tomography; and (III) {read-out} of the photon-number distribution. Because the time span of several operations varies widely, real scales are not used.

\section{Control of quenching dynamics}
\subsection{Time-dependence of the normalized coupling parameter $\xi$}
\begin{figure}
\centering
\includegraphics[width=14cm]{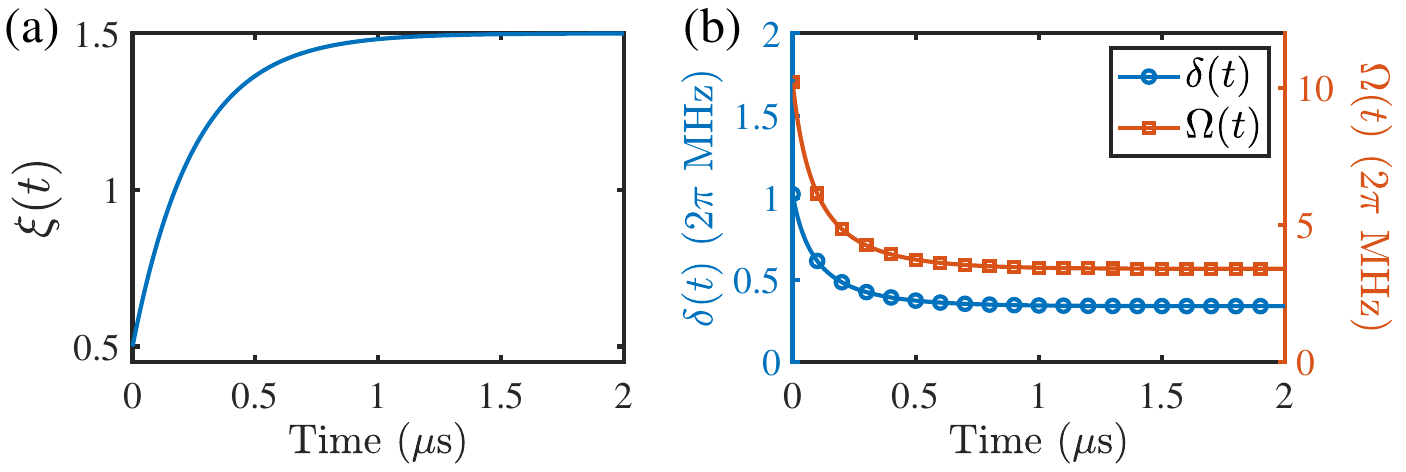}
\caption{(a) Normalized coupling parameter $\xi(t)$ versus time. 
	(b) Effective frequencies $\delta(t)$ and $\Omega(t)$ of the resonator and qubit versus time,
	showing on the left and right $y$-axes, respectively.}
\label{dwcdwq}
\end{figure}

During the quenching process, the normalized coupling parameter is changed as \cite{duanarxiv}

\begin{align} 
\xi(t)=\left(\xi_{\rm max}-\xi_{0}\right)\left[1- \exp\left({-\frac{8t}{t_f}}\right)\right]+\xi_{0}.
\end{align} 
Here, $\xi_{\rm max}$ ($\xi_{0}$) is the maximum (initial) value of $\xi(t)$ and $t_{f}$ is the total evolution time.
For Figs. 2 and 3 in the main text, we choose $\xi_{\rm max}=1.5$, $\xi_{0}=0.5$, and $t_f=2$ $\mu$s.  
In addition, we choose $\Omega=10\delta$ at all times of the evolution to ensure the preset limitation $\Omega \gg \delta$ of the SPT \cite{2015PRL}. 
The induced $\xi(t)$, $\Omega(t)$, and $\delta(t)$ are shown in Fig. \ref{dwcdwq}.

The experiment starts by tuning the test qubit to the operating frequency $\omega _{0}/2\pi= 5.18$ GHz, around which the two sine modulations are applied, with the fixed modulation frequencies $\nu _{1}/2\pi= 200$ MHz and $\nu _{2}/2\pi= 33.28$ MHz. 
The amplitude of the first modulation is set to $\varepsilon _{1}/2\pi= 165.85$ MHz, while $\varepsilon _{2}$ is taken as a control parameter. 
Near the operating frequency, the qubit is driven by a continuous microwave {with} Rabi frequency $K/2\pi= 19.91$ MHz.
For these settings, the dynamics of the test qubit and the resonator {are} governed by the Rabi Hamiltonian, with the qubit working in the {frame} rotating at the angular frequency $B_0$ relative to the laboratory frame. 
The resulting effective qubit-resonator coupling strength is $\eta/2\pi = 0.81$ MHz. 
During the Rabi dynamics, the ancilla qubit $Q_{2}$ is far off-resonant with the test qubit $Q_{1}$ and with the resonator, so that it remains in the ground state.


\subsection{Control of the effective qubit frequency $\Omega$}
\begin{figure}
\centering
\includegraphics[width=18cm]{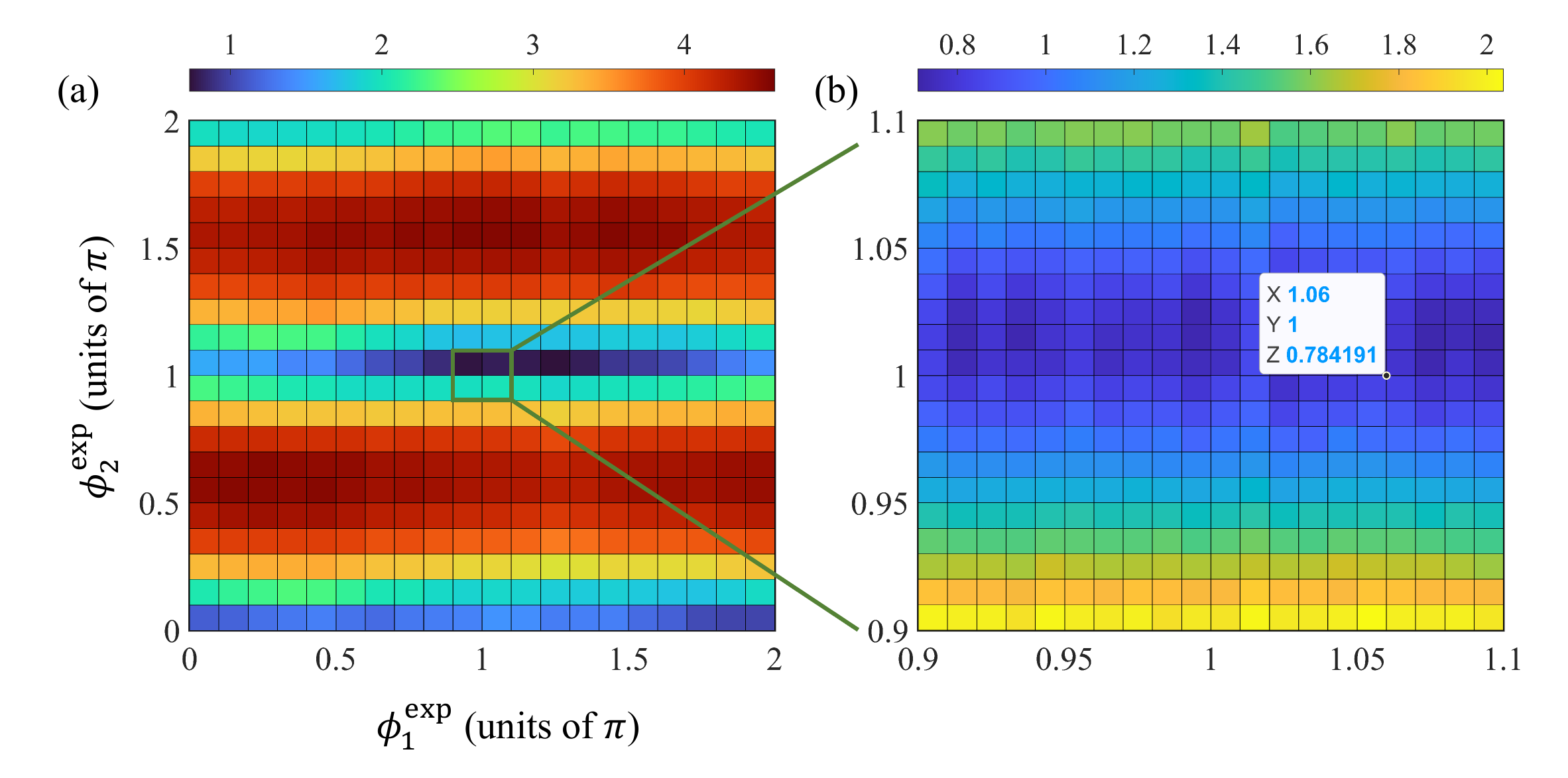}
\caption{(a) Fitting errors (Euclidean norm) versus phase modifications $(\phi_1^{\rm exp}$ and $\phi_2^{\rm exp})$ in the experiment. (b) Magnified view of (a) when  $\{\phi_1^{\rm exp},\phi_2^{\rm exp}\}=[0.9,1.1]\pi$, where the optimal point used in the experiment has been marked.}
\label{fitp0p1}
\end{figure}

\begin{figure}
\centering
\includegraphics[width=16cm]{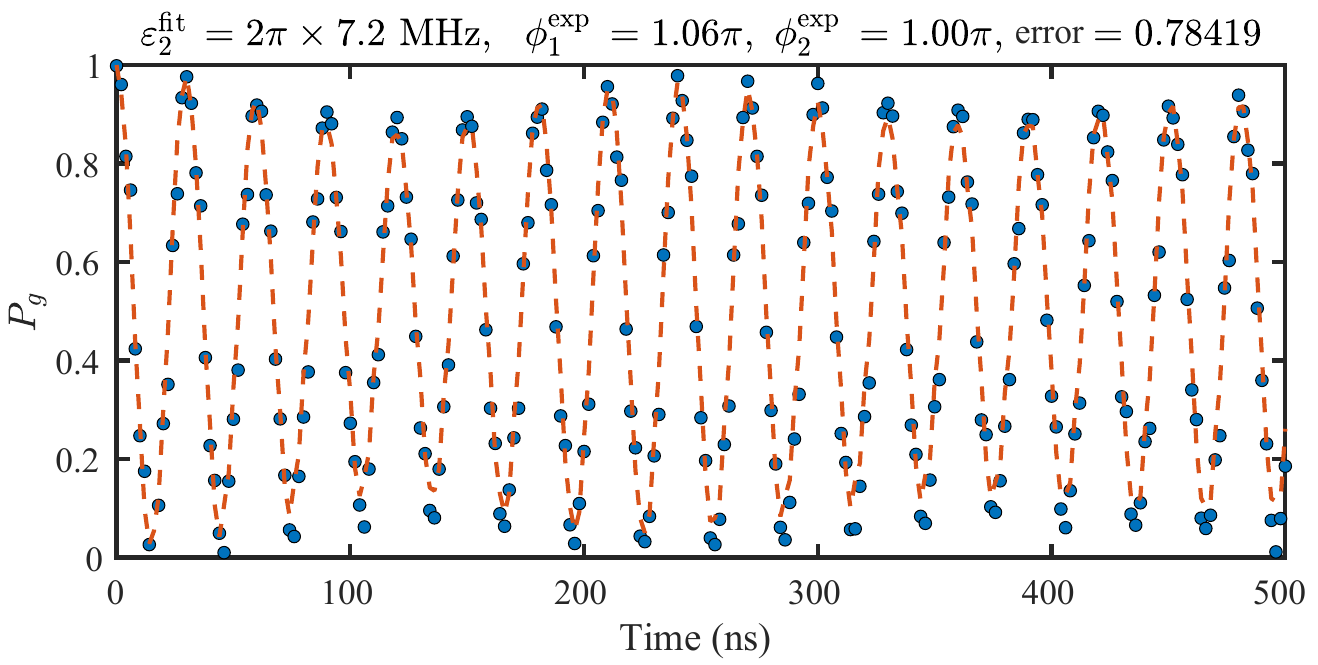}
\caption{ Fitting for $\phi_{1}^{\rm{exp}}=1.06\pi$ and $\phi_{2}^{\rm{exp}}=1.00\pi$. The populations $P_g$ in the ground state of the test qubit $Q_{1}$ versus time. The experiment data and the numerical results are marked by dots and a dashed curve, respectively.}
\label{fitsituation}
\end{figure}

Because of the imperfect waveform of the periodically modulated excitation energy $\hbar \omega_q(t)$, 
{we modify the first and second modulating pulses as $\varepsilon_1\cos (\nu_1 t+\phi_1^{\rm exp}) $ and $\varepsilon_2\cos (\nu_2 t+\phi_2^{\rm exp})$, respectively, to optimize the most appropriate dynamics of the Rabi Hamiltonian $H_{R}$.}
We iterate over different phases $(\phi_1^{\rm exp}, \phi_2^{\rm exp})$ to carry out the experiments with setting $\delta=0$,
and finally {obtain} the one which best coincides with the corresponding simulated Rabi oscillation curves. The optimal results, specifically, for $\Omega/2\pi= 3.6$ MHz, are shown in Fig. \ref{fitp0p1}, where we can see that the  fitting error (the Euclidean norm) becomes minimal when $\phi_1^{\rm exp}=1.06\pi$ and $\phi_2^{\rm exp}=1.00\pi$.
In this case, the concrete 
fitting situation is indicated in Fig. \ref{fitsituation}. The experimental data are intuitively close to the numerical fitting in Fig. \ref{fitsituation}, confirming the validity of such phase modification.


For convenience, we choose the same phase modifications (i.e., $\phi_{1}^{\rm{exp}}=1.06\pi$ and $\phi_{2}^{\rm{exp}}=1.00\pi$) for different $\Omega$ throughout the experiments. Note that we also slightly adjust the center frequencies and amplitudes of the two modulating pulses to improve the control.

\subsection{Control of the effective resonator frequency $\delta$}
\begin{figure}[htb]
\centering
\includegraphics[width=18cm]{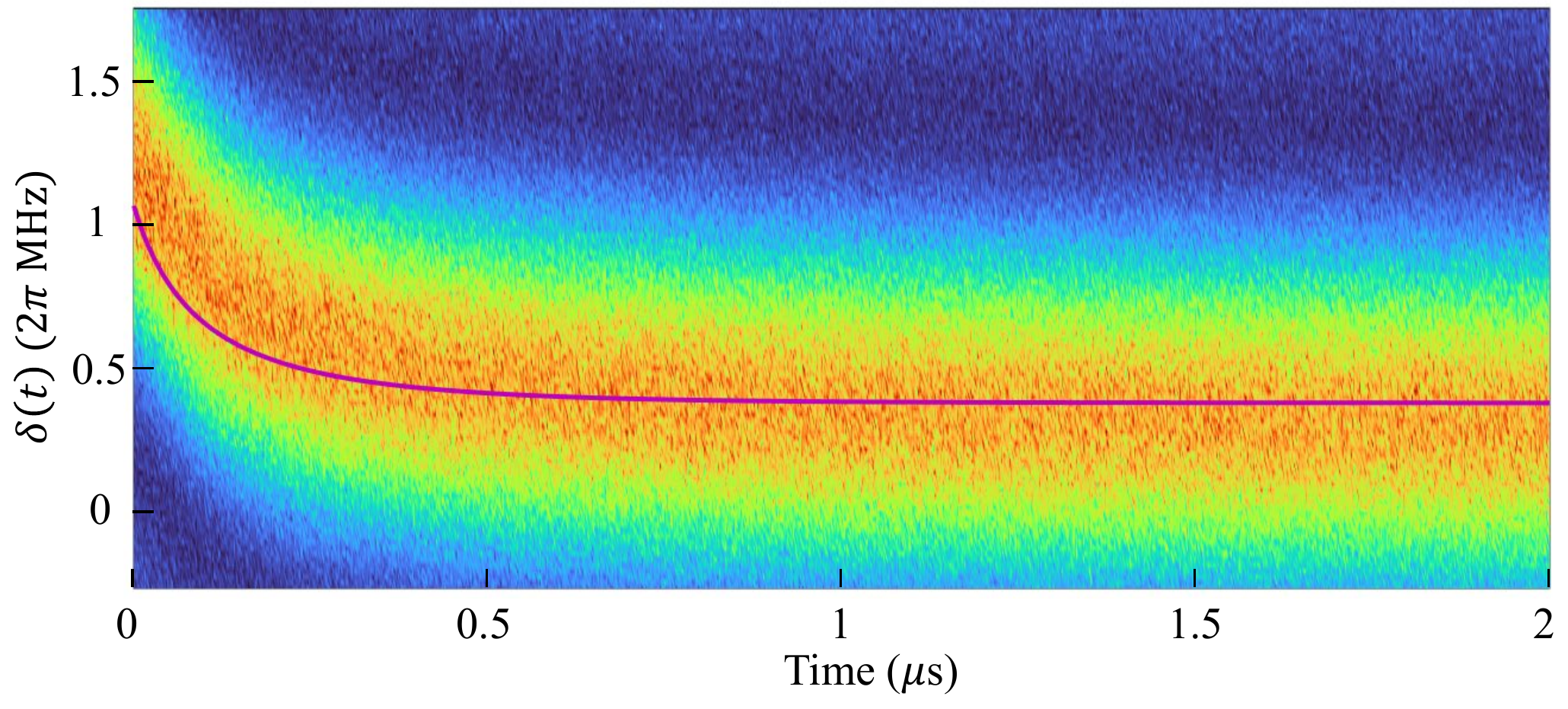}
\caption{Experimental frequency $\delta(t)$ of the resonator versus time, induced by tuning the frequency of the ancilla qubit $Q_{2}$ with a suitable translation added to the $y$-axis. The ideal $\delta(t)$ is plotted with a solid purple-solid curve.}
\label{dwc}
\end{figure}

The control of the effective resonator frequency $\delta(t)$ of the effective Rabi Hamiltonian in Eq.~(\ref{eqS9}) can be 
achieved by adjusting the Stark shift induced by the ancilla qubit $Q_{2}$. We can control this Stark shift by tuning the frequency
of the ancilla qubit. When the detuning $\Delta\omega$ between this ancilla qubit and
the resonator is varied from $\Delta \omega $ to $\Delta \omega^{\prime }$, and remains much larger than their coupling strength $\lambda' $, the resulting
resonator frequency shift {becomes} 

\begin{eqnarray} \label{omega}
\delta \omega _{S}=
\frac{\lambda'^{2}}{\Delta \omega}-\frac{\lambda'^{2}}{\Delta \omega^{\prime }},
\end{eqnarray}
where $\lambda'/2\pi=20.91$ MHz (different from $\lambda$ in the main text).
Equation (\ref{omega}), in a different form, reads

\begin{eqnarray} \label{omega2}
\delta(t)-\delta(0)=\frac{\lambda'^{2}}{f(0)-\omega_p}-\frac{\lambda'^{2}}{f(t)-\omega_p},
\end{eqnarray}
with $f(t)$ being the transition frequency of the ancilla qubit. Based on Eq. (\ref{omega2}), we change $f(t)$ from the idle frequency $f_{\rm idle}$, viz., $f(0)/2\pi=f_{\rm idle}/2\pi= 5.93$ GHz, yielding,

\begin{eqnarray} \label{ft}
f(t)=\omega_p-\left[\frac{\delta(t)-\delta(0)}{\lambda'^2}-\frac{1}{f_{\rm idle}-\omega_p}\right]^{-1}.
\end{eqnarray}
To demonstrate the effectiveness of controlling $\delta(t)$ experimentally, we apply square pulses with amplitudes $f(t)$ 
($x$-axis) to the Z-line of the ancilla qubit $Q_{2}$. 
Meanwhile, several square-envelope pulses with certain amplitudes and specific 
frequencies $\omega_p+\delta(t)$ ($y$-axis) are applied to the XY-line of the test qubit $Q_{1}$ to excite the resonator (by cross-talk interactions). 
Subsequently, we measure {the} populations of the ancilla qubit $Q_{2}$ ($z$-axis) after a qubit-resonator-swap interaction (span time $\pi/\lambda'$).  
This spectroscopy reflects the relationship between applying $f(t)$ and the induced offset of the effective resonator frequency $\delta(t)$. 
After appropriately translating the $y$-axis and remapping values of $f(t)$ to the corresponding time,
we achieve the experimental control of $\delta(t)$ as shown in Fig. \ref{dwc}, where the high-value populations (highlighted area) roughly depict the trend of the experimental  $\delta(t)$ and coincide well with its ideal values.

\section{Effects of off-resonant couplings and decoherence}\label{Stark}
When considering high-energy levels of the Xmon qubit, the Hamiltonian becomes

\begin{eqnarray}
H&=&H_{0}+H_{I}, \\
H_{0} &=&(\omega _{0}+2\nu _{1})a^{\dagger }a+[\omega _{0}+\varepsilon
_{1}\cos (\nu _{1}t)]q^{\dagger }q-\frac{\gamma}{2}q^{\dagger 2}q^{2}, \\
H_{I} &=&\delta  a^{\dagger }a+(\lambda a^{\dagger }q+{\rm{h.c.}}),
\end{eqnarray}%
where $q$ ($q^\dagger$) denotes the annihilation (creation) operator
for the Xmon qubit mode and $\gamma$ is the anharmonicity of the qubit. For simplicity, 
we ignore  the microwave driving $K$ and the frequency modulation $\varepsilon_{2}$. Performing the transformation $U_{0}$ and
considering the lowest three levels of the Xmon qubit, \{$|g\rangle$, $|e\rangle$, $|f\rangle$\}, we obtain the system Hamiltonian in the rotating frame as

\begin{eqnarray}
H_{I}^{\prime } &=&\delta a^{\dagger }a+\left\{\exp \left[ -i\mu \sin (\nu _{1}t)\right] \lambda \exp
\left( 2i\nu _{1}t\right) a^{\dagger }[\left\vert g\right\rangle
\left\langle e\right\vert +\sqrt{2}\exp \left( i\gamma t\right) \left\vert
e\right\rangle \left\langle f\right\vert ]+{\rm{h.c.}}\right\}.
\end{eqnarray}%
Using the Jacobi-Anger
expansion, we obtain%

\begin{eqnarray}
H_{I}^{\prime } &=&\delta a^{\dag}a+\left(\stackrel{\infty }{%
	\mathrel{\mathop{\sum }\limits_{n=-\infty }}%
}J_{n}(\mu )\lambda a^{\dagger }\left\{\exp \left[-i(n-2)\nu _{1}t\right]\left\vert
g\right\rangle \left\langle e\right\vert +\sqrt{2}\exp [-i(n-2)\nu
_{1}t+i\gamma t]\left\vert e\right\rangle \left\langle f\right\vert \right\}+{\rm{h.c.}}\right).
\end{eqnarray}%
By assuming that \{$\left\vert \lambda J_{0}(\mu )\right\vert, \ \delta \} \ll \nu _{1}$, $H_{I}^{\prime }$ reduces to the effective Hamiltonian

\begin{eqnarray}
H_{e} &=&\left[J_{2}(\mu )\lambda a^{\dagger }\right]\left\vert g\right\rangle
\left\langle e\right\vert +{\rm{h.c.}} \cr\cr
&&+\;S_{1}(\left\vert g\right\rangle \left\langle g\right\vert -\left\vert
e\right\rangle \left\langle e\right\vert )a^{\dagger }a-S_{1}\left\vert
e\right\rangle \left\langle e\right\vert +S_{2}\left\vert e\right\rangle
\left\langle e\right\vert a^{\dagger }a+ \delta a^{\dagger }a,
\end{eqnarray}%
where 

\begin{eqnarray}
S_{1} &\simeq &\frac{[J_{0}(\mu )\lambda ]^{2}}{2\nu _{1}}+\frac{[J_{1}(\mu
	)\lambda ]^{2}}{\nu _{1}}+\frac{[J_{-1}(\mu )\lambda ]^{2}}{3\nu _{1}}, \\ \cr
S_{2} &\simeq &\frac{2[J_{0}(\mu )\lambda ]^{2}}{2\nu _{1}+\gamma }+\frac{%
	2[J_{1}(\mu )\lambda ]^{2}}{\nu _{1}+\gamma }+\frac{2[J_{-1}(\mu )\lambda ]^{2}%
}{3\nu _{1}+\gamma }.
\end{eqnarray}%
The term $S_{2}\left\vert e\right\rangle \left\langle e\right\vert
a^{\dagger }a$ results from a dispersive coupling to the second-excited
state $\left\vert f\right\rangle $. 

\begin{figure}[b]
\centering
\includegraphics[width=15cm]{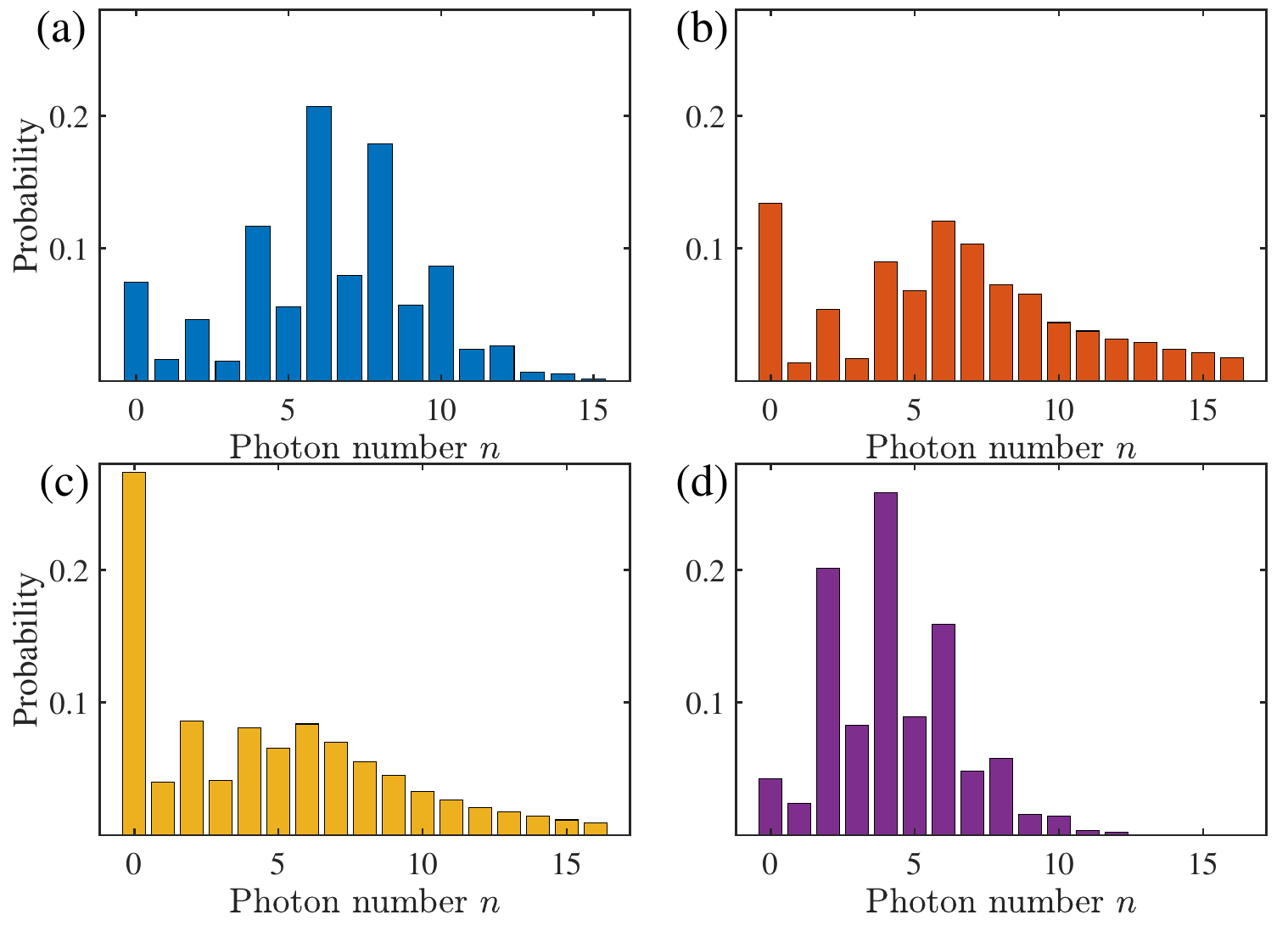}
\caption{Numerically simulated photon-number distributions of fields at $t=2$ $\mu$s,
which are  calculated by the
(a) effective Hamiltonian without decoherence,
(b) full Hamiltonian without decoherence,
and (c) full Hamiltonian with decoherence.
{(d) Ground state results of the effective Hamiltonian.}}
\label{phocompar}
\end{figure}

Discarding the constant term, we can
rewrite $H_{e}$ as%

\begin{eqnarray}
H_{e}=H_{e,1}+H_{e,2},
\end{eqnarray}
where%

\begin{eqnarray}
H_{e,1}=\left[J_{2}(\mu )\lambda a^{\dagger }\right]\left\vert g\right\rangle
\left\langle e\right\vert +{\rm{h.c.}},
\end{eqnarray}
\begin{eqnarray}
H_{e,2}=\left(S_{1}-\frac{1}{2}S_{2}\right)\left(\left\vert g\right\rangle \left\langle
g\right\vert -\left\vert e\right\rangle \left\langle e\right\vert
\right)a^{\dagger }a+\frac{1}{2}S_{1}(\left\vert g\right\rangle \left\langle
g\right\vert -\left\vert e\right\rangle \left\langle e\right\vert )+\left(\frac{1}{2}S_{2}+\delta\right)a^{\dagger }a.
\end{eqnarray}
Therefore, the Stark shifts are significantly reduced when the
effect of the second-excited
state $\left\vert f\right\rangle $ is considered. For the present device parameter
setting, we obtain 

\begin{eqnarray}
S_{1}-\frac{1}{2}S_{2}\ \sim\ 2\pi \times 0.45 ~{\rm{MHz}}.
\end{eqnarray} 
When the
microwave drive is applied, $H_{e,2}$ becomes

\begin{eqnarray} \label{29}
H_{e,2}^{\prime }=\frac{\left(S_{1}-\frac{1}{2}S_{2}\right)^{2}}{2K} \sigma _{x}\left(a^{\dagger }a\right)^{2}+\frac{\left(\frac{1}{2}S_{1}\right)^{2}}{2K}\sigma _{x}+\left(\frac{1}{2}S_{2}+\delta\right)a^{\dagger}a,
\end{eqnarray}
where 
\begin{eqnarray}
\frac{\left(S_{1}-\frac{1}{2}S_{2}\right)^{2}}{2K}\simeq \frac{2\pi \times
	0.45^{2}}{40} \ {\rm MHz}=2\pi \times 0.0061 \ {\rm MHz}.
\end{eqnarray}
%

The first term of Eq. (\ref{29}) produces a qubit-state-dependent Kerr effect on the photonic field, dispersing the phase-space distributions of the coherent fields, which partly accounts for distortions of the observed Gaussian wavepackets. 
To quantitatively explore influences arising from imperfect Hamiltonian dynamics and decoherence, we perform numerical simulations on the photon-number distributions at $t=2$ $\mu$s. 
Figures \ref{phocompar}(a) and \ref{phocompar}(b) present results based on the effective Rabi Hamiltonian and the full Hamiltonian{, respectively; while} Fig. \ref{phocompar}(c) displays the result calculated by the master equation, including both the full Hamiltonian dynamics and decoherence. 
{Figure \ref{phocompar}(d) shows the photon number distributions associated with the ground state of the effective Rabi Hamiltonian.}
{For the ground state of the ideal Rabi model with an infinite frequency ratio \cite{2015PRL}, the vacuum component has a negligible population. Due to the experimental imperfections, the observed output state has a significant vacuum population of about 0.29. To quantify individual contributions to the vacuum population, we perform numerical simulations step by step, including more and more experimental imperfections. Thus obtained vacuum populations are respectively 0.07, 0.13, 0.27, and 0.04, as shown in Figs. \ref{phocompar}(a)-(d). The results imply that the limitation of the effective qubit-resonator frequency ratio, non-adiabaticity, deviation from model Hamiltonian, and dissipation contribute vacuum populations of 0.04, $0.07-0.04=0.03$, $0.13-0.07=0.06$, and $0.27-0.13=0.14$, respectively. We note the calculated vacuum population 0.27 in Fig. \ref{phocompar}(c) is slightly smaller than the observed result 0.29 in Fig. 2 in the main text, mainly due to deviations of the system parameters used for the simulation from their real values.}
In the simulation, the dissipation times for the qubit and the resonator are the same as those listed in Table \ref{parameter}.
These results clearly show {that} the population of the vacuum state is mainly caused by decoherence.

\begin{figure}
\centering
\includegraphics[width=10cm]{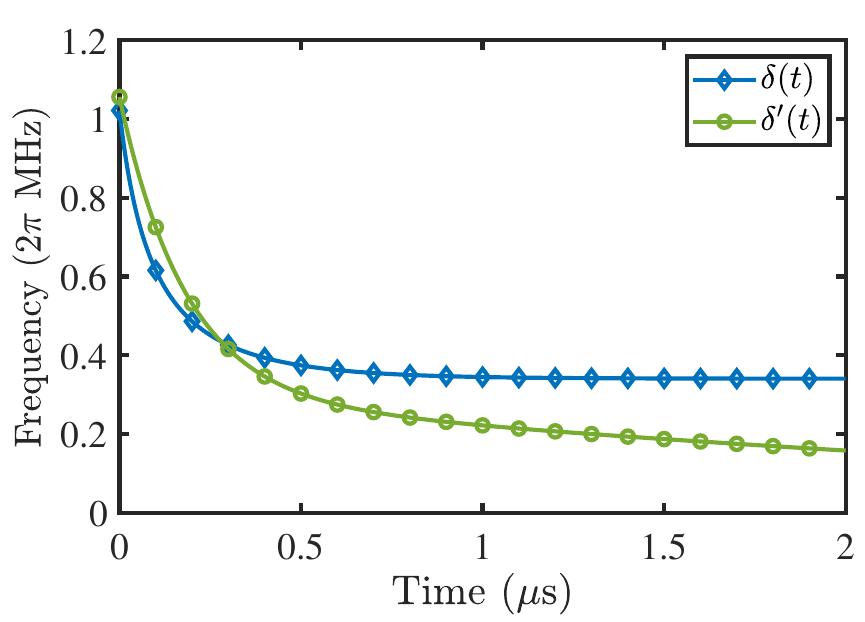}
\caption{Effective resonator frequency versus time. 
The green line with circular markers represents the fitting effective resonator frequency $\delta'(t)$ deduced by experimental measurements in Fig. \ref{dwc}. 
The blue line with diamond markers indicates the ideal effective resonator frequency.}
\label{dwc0&dwc}
\end{figure}

Note {that} there are also two corrections in the numerical simulation for the master equation:
(i) applying $\Omega(t)'=1.56$~$\Omega(t)$ [$\Omega(t)$ shown in Fig. \ref{dwcdwq}];
(2) utilizing the fitting $\delta'(t)$ deduced by experimental measurements in Fig. \ref{dwc}, shown in Fig. \ref{dwc0&dwc}.

\begin{figure}[b]
\centering
	\includegraphics[width=14.5cm]{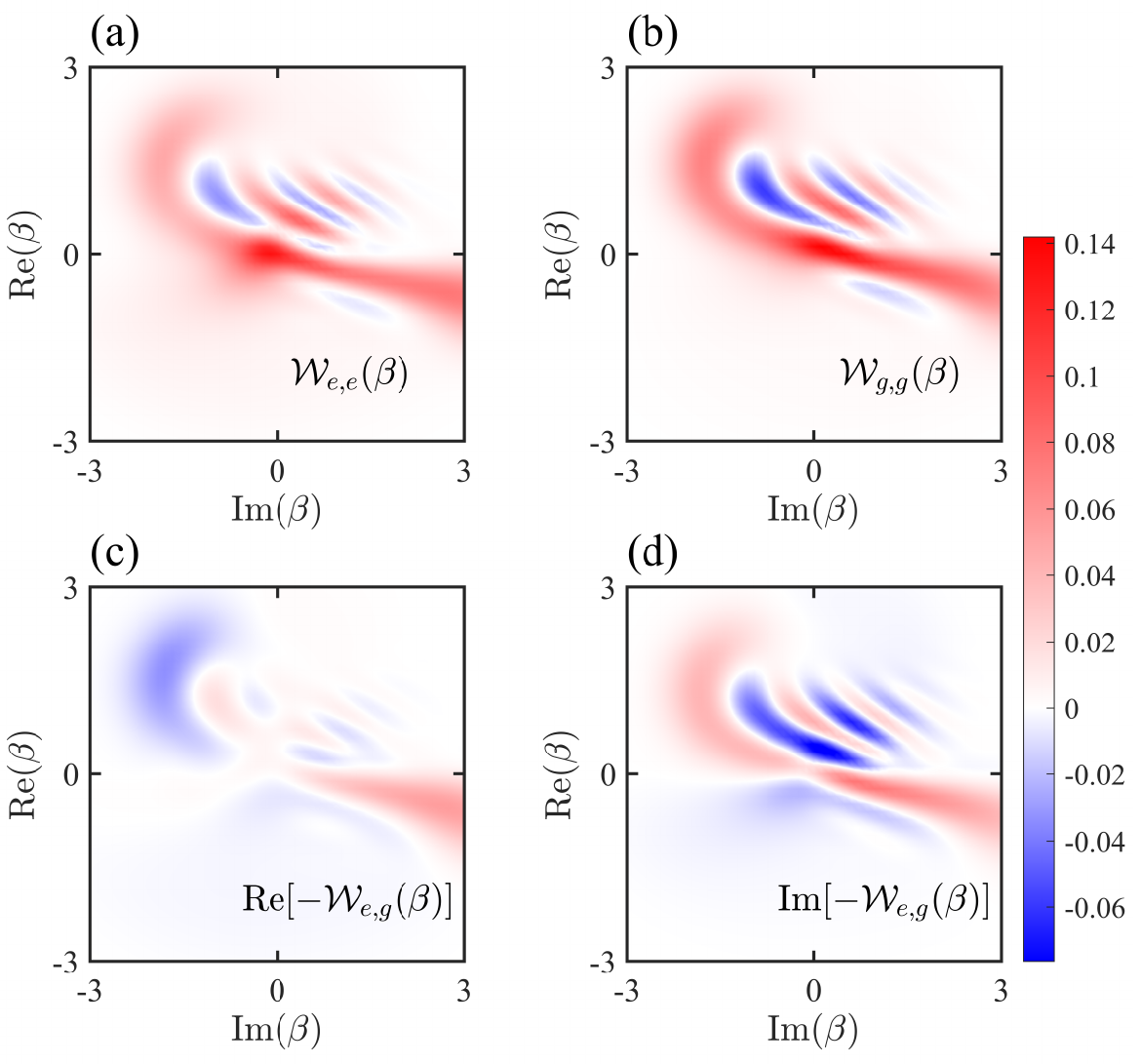}
\caption{Numerically simulated Wigner matrix tomography. 
	(a--d) Corresponds to Fig. 3 (a--d) in the main text, respectively.
	Note that suitable rotations of Wigner tomography have been applied and all the data are measured at $t=2~\mu$s based on the parameter corrections in Sec.~\ref{Stark}.}
\label{simwig}
\end{figure}

\begin{figure}[h]
\centering
\includegraphics[width=10cm]{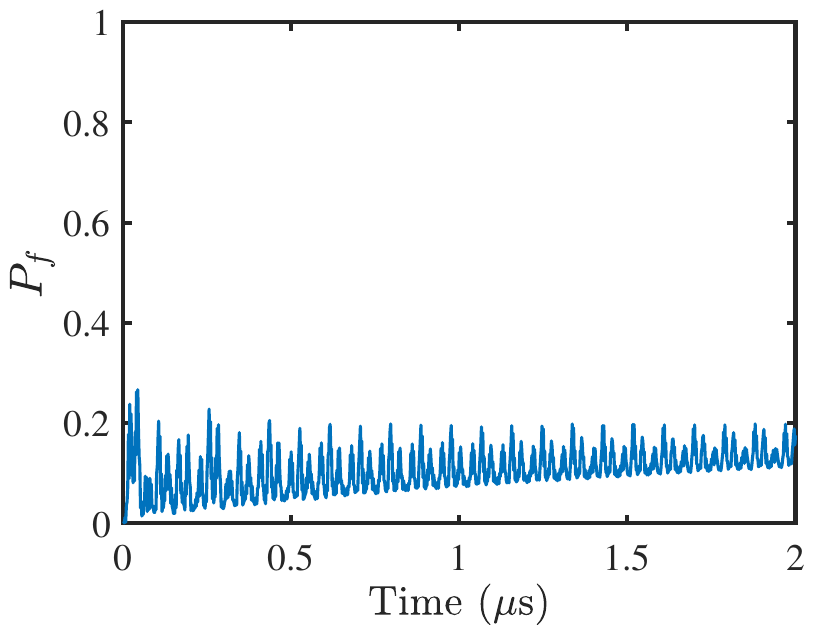}
\caption{Numerically simulated population of the third level $|f\rangle$. }
\label{Pf}
\end{figure}

The simulated average photon number has shown in Fig. 2 in the main text. 
Additionally, based on such numerical simulation, 
we plot the corresponding Wigner functions in Fig. \ref{simwig}.
We also plot the population of the third level $|f\rangle$ in Fig. \ref{Pf}, when the $|f\rangle$ is included in the numerical simulation. The results show that the average population of the third level is about 0.11 during the quenching dynamics.

%
%

%

\section{Characterization of the qubit-resonator state}
\subsection{Photon-number distribution}\label{ss}
\begin{figure}[htb]
\centering
\includegraphics[width=16cm]{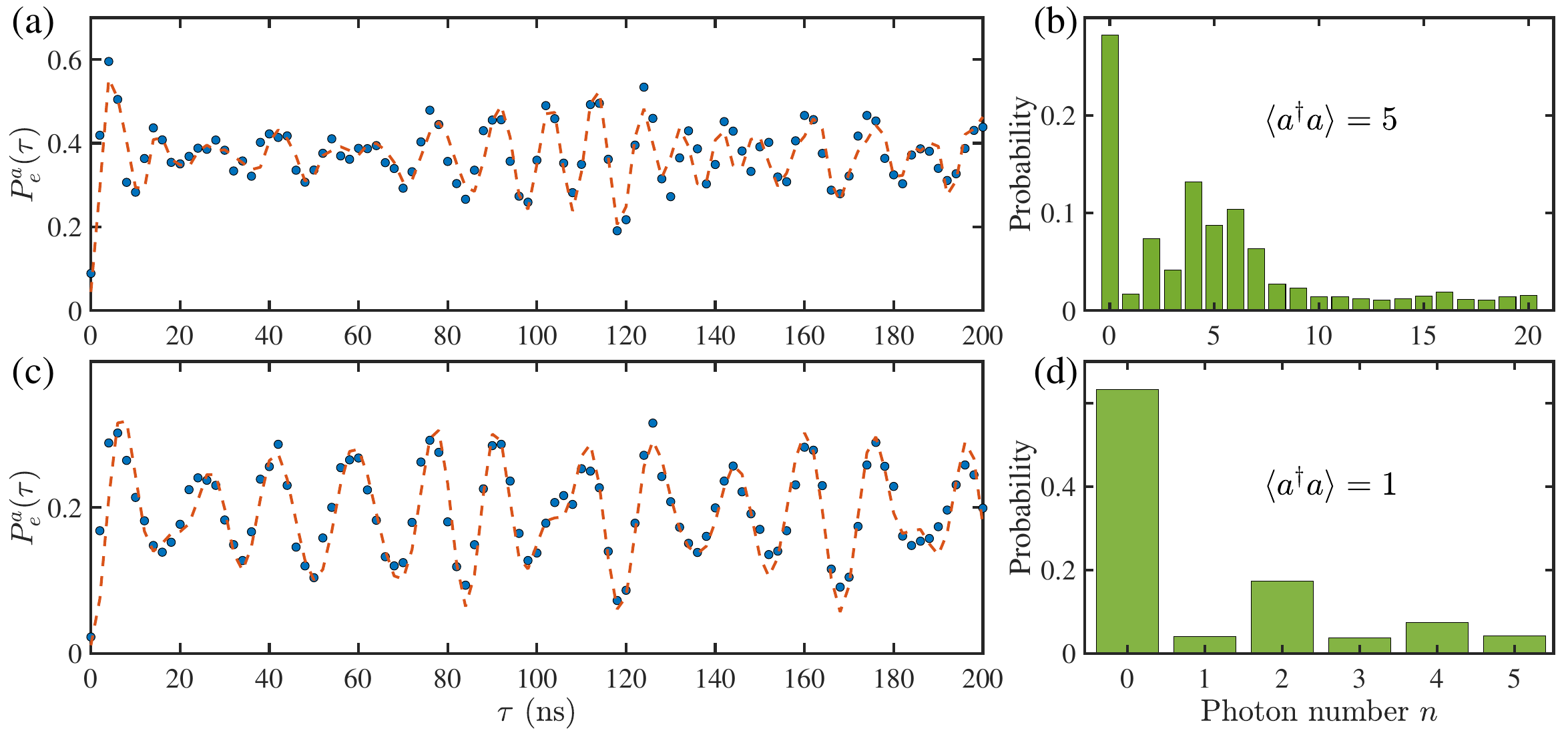}
\caption{Experimental (dots) and theoretical (dashed curves) time-resolved quantum Rabi oscillations and corresponding photon-number distributions
for (a, b) $\langle a^\dag a \rangle=5$ and (c, d) $\langle a^\dag a \rangle=1$.}
\label{photondis}
\end{figure}

All the measured average photon numbers and the Wigner function values in the main text are deduced from the photon-number distribution. In the experiment, 
after carrying out the part of the SPT (or after  Wigner   tomography), the microwave drive $K$ and the frequency modulations $\omega_{q}$ are switched off. 
Meanwhile, the ancilla qubit $Q_{2}$ is tuned on resonance with the resonator (frequency $5.581$ GHz) from the idle frequency $5.93$ GHz. 
Furthermore, the ancilla qubit $Q_{2}$ undergoes photon-number-dependent Rabi oscillations. 
The populations $P_{e}^a(\tau)$, of the excited state of the ancilla qubit $Q_{2}$ for a given interaction time $\tau$, are measured by biasing the ancilla back to its idle frequency, 
where its state is read out  (intuitively see Fig. \ref{sequence}). The recorded time-resolved quantum Rabi oscillation signals can be fitted as

\begin{eqnarray}\label{Pnfit}
P_{e}^{a}(\tau)=\frac{1}{2}\left[1-P_g^a(0)\stackrel{n_{\max }}{%
\mathrel{\mathop{\sum }\limits_{n=0}}%
}P_{n}e^{-\kappa _{n}\tau}\cos \left(2\sqrt{n}\lambda' \tau\right)\right],
\end{eqnarray} 
where $P_{n}$ denotes the photon-number distribution probability, $P_g^a(0)$ indicates the probability for the ancilla qubit $Q_{2}$ to 
start in the ground state, $n_{\max} $ is the cutoff of the photon number,  and $\kappa_{n}=n^{l}/T_{1,p}$ ($l=0.7$) \cite{Pnfit1,Pnfit2,Pnfit3,Pnfit4,Pnfit5} 
is the empirical decay rate of the Rabi oscillations associated with the $n$-photon state. It is worth mentioning that, when there are a lot of photons 
in the resonator, especially $\langle a^{\dag}a\rangle>10$, the detuning $\Delta\omega/2\pi=(5.93-5.581) \ {\rm GHz}=0.329$ GHz is not large enough to avoid interactions between 
the ancilla qubit $Q_{2}$ and the resonator. This induces a minor excitation of the ancilla qubit $Q_{2}$, depending on the excitation of the resonator. 
Thus, referring to \cite{Pnfit1}, we ignore the {small} excitation {of $|e\rangle$} and rescale the factor $1/2$ in the photon-number-dependent Rabi oscillations equation \cite{Pnfit5}, to $P_g(0)/2$ [see Eq. (\ref{Pnfit})]. Based on the measurements and fitting operations described above, 
we can give the fitting {situation} and corresponding photon-number distribution in Fig. \ref{photondis}.

\subsection{Diagonal Wigner matrix elements}\label{pro}
As described in the main text, the Wigner function {is given by}

\begin{eqnarray}
{\cal W}_{k,k'}(\beta )=\frac{2}{\pi }\stackrel{\infty }{\mathrel{\mathop{\sum }\limits_{n=0}}}(-1)^{n}{\cal P}_{n}^{k,k'}(\beta ), 
\end{eqnarray}
where 

$${\cal P}_{n}^{k,k'}(\beta )=\left\langle n\right\vert D(-\beta )\rho_{k,k'}D(\beta )\left\vert n\right\rangle. \ \ \ \ \ \ \ \  (k,k'=e,g)$$ 
To measure the Wigner diagonal elements, a displacement operation, $D(\beta )=\exp \left(\beta a^{\dagger }-\beta^{*}a\right)$, is applied to the resonator, following which the ancilla qubit $Q_{2}$ is resonantly coupled to the resonator for a given time $\tau$.
Then, the ancilla qubit $Q_{2}$ is biased to the idle frequency for the {read-out} of states. 

The photon-number distributions ${\cal P}_{n}^{g,g}(\beta )$ and ${\cal P}_{n}^{e,e}(\beta )$ of the displaced light field are associated with the test qubit states $\left\vert g\right\rangle $ and $\left\vert e\right\rangle$.
Such distributions can be extracted from the excited-state populations [$P_{n,e}^{g}(\beta ,\tau)$ and $P_{n,e}^{e}(\beta ,\tau)$] of the ancilla qubit, conditional on the detection of the test qubit in
states $\left\vert g \right\rangle $ and $\left\vert e\right\rangle $. Thus, the normalized Wigner function (for diagonal elements in the qubit basis) is given by

\begin{eqnarray}
W_{k,k}(\beta)={\cal W}_{k,k}(\beta )/P_k,
\end{eqnarray} 
with $P_k$ being the $\left\vert k\right\rangle $-state population of the test qubit. To completely show the distribution of the generated cat states in phase space, we calibrate
$\{{\rm Re}(\beta),{\rm Im}(\beta)\}\in[-3,3]$, implying that the displacement distance of $D(-\beta )$, i.e., $|\beta|$, can be up to $3\sqrt{2}$. 
This leads to the fact that the displaced light fields in some cases process an average photon number $\langle a^\dag a \rangle\gtrsim18$. 

\begin{figure}[htb]
\centering
\includegraphics[width=15cm]{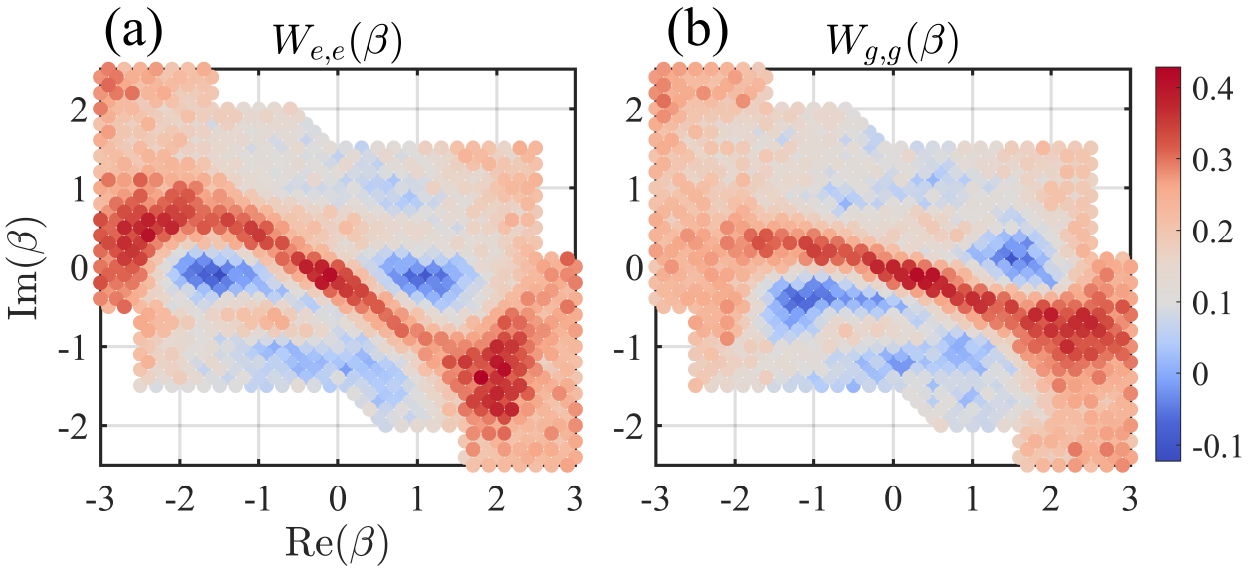}
\caption{ Experimental data of the Wigner functions (a) $W_{e,e}(\beta)$ and (b) $W_{g,g}(\beta)$, without displaying the points corresponding to large numbers of photons ($\langle a^\dag a \rangle>10$).}
\label{expwigdata}
\end{figure}

However, as claimed in Sec. \ref{ss}, the {read-out} of the photon-number distribution becomes imprecise with photons growing. We therefore ignore the areas with
large numbers of photons and utilize the remaining areas to deduce the normalized density matrix $\rho_{k,k}/P_k$ of light fields. This {treatment} is intuitively shown in Fig. \ref{expwigdata}. The calculation from the Wigner function $W_{k,k}(\beta)$ to the density matrix $\rho_{k,k}/P_k$ is completed {using} convex optimization, supported by the CVX toolbox based on MATLAB \cite{cvx}. The solved density matrix $\rho_{k,k}/P_k$ is Hermitian and positive semidefinite as well as satisfying ${\rm Tr}(\rho_{k,k}/P_k)=1$. Furthermore, we use this solved $\rho_{k,k}/P_k$ to plot ${\cal W}_{k,k}(\beta)$, shown in Figs. 3(a) and 3(b) in the main text, which has the same distribution in phase space as Figs. \ref{expwigdata}(a) and \ref{expwigdata}(b) here. As for Figs. 3(c) and 3(d) in the main text, we additionally perform Wigner tomography (detailed in the next section) and take the same {treatment} as that for Figs. \ref{expwigdata}(a) and \ref{expwigdata}(b). 
The time to measure the Wigner function is $t=$1.946 $\mu$s because the error of {the} qubit projection is relatively small at this time.

\subsection{Off-diagonal Wigner matrix elements}
To fully characterized the nonclassical light-matter correlations, it is necessary to reconstruct  the off-diagonal elements,
measurements of which require Wigner tomography in the rotated basis $\left\{ \left\vert \pm _x\right\rangle =(\left\vert e\right\rangle \pm \left\vert g\right\rangle /\sqrt{2}\right\} $ and $\left\{ \left\vert \pm_y\right\rangle =(\left\vert e\right\rangle \pm i\left\vert g\right\rangle /\sqrt{2}\right\} ${, respectively}. The Wigner diagonal elements in these bases are related to both diagonal and off-diagonal elements in the basis $\left\{ \left\vert e\right\rangle ,\left\vert g\right\rangle \right\} $:

\begin{eqnarray}
{\cal W}_{\pm x,\pm x}(\beta ) &=&\frac{1}{2}[{\cal W}_{e,e}(\beta )+{\cal W}_{g,g}(\beta )]\pm \mathop{\rm Re}[{\cal W}_{e,g}(\beta )], \\
{\cal W}_{\pm y,\pm y}(\beta ) &=&\frac{1}{2}[{\cal W}_{e,e}(\beta )+{\cal W}_{g,g}(\beta )]\mp \mathop{\rm Im}[{\cal W}_{e,g}(\beta )].
\end{eqnarray}%
To measure the Wigner matrix element ${\cal W}_{+x,+x}(\beta )$, a rotation $R_{y}(\pi /2)$ is performed on the test qubit $Q_{1}$ before its state {read-out},
transforming the $x$-axis basis states $\left\vert +_x\right\rangle $ and $\left\vert -_x\right\rangle $ to the $z$-axis basis states $\left\vert e\right\rangle $ and $|g\rangle${, respectively}, which can be directly measured by the {read-out} resonator. After this rotation and the 
resonator displacement $D(-\beta )$, the measured photon-number distribution ${\cal P}_{n}^{e,e}(\beta )$, 
conditional on the detection of the test qubit state $\left\vert e\right\rangle $, 
yields the conditional Wigner function $W_{+x,+x}(\beta )$. 
Similarly, performing the resonator displacement $D(-\beta)$ after
a qubit rotation $R_{x}(\pi/2)$, we can reconstruct the conditional Wigner function $W_{+_{y},+_{y}}$
based on the measured photon-number distribution $\mathcal{P}_{n}^{e,e}(\beta)$.
The elements ${\cal W}_{\pm _J,\pm _J}(\beta )$ ($J=x,y$) 
are related to the normalized Wigner function $W_{\pm _J,\pm _J}(\beta )$ by 

\begin{eqnarray}
	{\cal W}_{\pm _J,\pm _J}(\beta )=P_{\pm _J}^{e(g)}W_{\pm _J,\pm _J}(\beta ),
\end{eqnarray}
where $P_{J}^{e(g)}$ denotes the $\left\vert (e)g\right\rangle $-state population of the test qubit $Q_{1}$ after the corresponding frame transformation.  
Moreover, the real and imaginary parts of the off-diagonal element ${\cal W}_{e,g}(\beta )$, can be calculated by

\begin{eqnarray}
{\rm Re}[{\cal W}_{e,g}(\beta)]&=&\frac{1}{2}[{\cal W}_{+x,+ x}(\beta )-{\cal W}_{-x,-x}(\beta )], \cr\cr
{\rm Im}[{\cal W}_{e,g}(\beta)]&=&\frac{1}{2}[{\cal W}_{-y,-y}(\beta )-{\cal W}_{+y,+y}(\beta )].
\end{eqnarray}
Note that ${\cal W}_{\pm_x,\pm_x}$ and ${\cal W}_{\pm_y,\pm_y}$ can be described the same as ${\cal W}_{k,k}$ in Sec. \ref{pro}. The images of ${\rm Re}[-{\cal W}_{e,g}(\beta)]$ and ${\rm Im}[{\cal W}_{e,g}(\beta)]$ are shown in Figs. 3(c) and 3(d) in the main text, respectively.

\begin{figure}[htb]
\centering
\includegraphics[width=18cm]{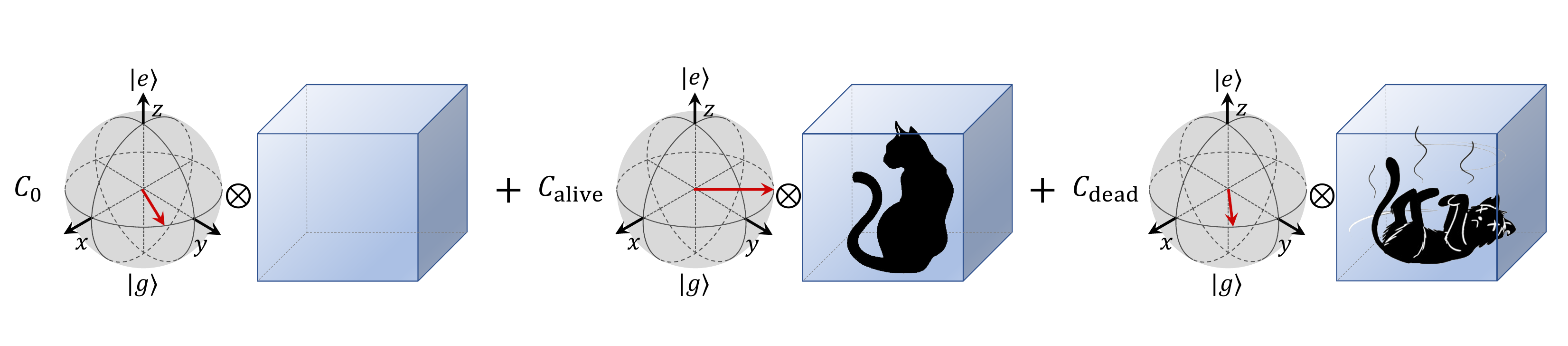}
\caption{Pictorial illustration of the three-component cat state. The resonator is in a superposition of being empty (without any cat) and containing a coherent field, which is formed by two superimposed components with complex amplitude $\alpha$ (alive cat) and $-\alpha$ (dead cat). These field components are correlated with different Bloch vectors (red-arrow lines) of the qubit.}
\label{EC}
\end{figure}

{\section{Measure of the qubit-resonator entanglement}
The partial transpose of the density matrix is 

\begin{eqnarray}
  	{\rho^{\Gamma_Q}} =\sum_{k=e,g}\sum_{k'=e,g}\rho_{k',k}\otimes|k\rangle\langle k'|,
\end{eqnarray}
whose eigenvalues are defined as $E_i$. The negativity \cite{nega} is the absolute sum of the negative eigenvalues, viz., $\mathcal{N}(\rho)=\left|\sum_{E_i<0}E_i\right|=0.2483$. For comparison, the upper bound of the negativity, is $\mathcal{N}(|\psi_{\rm sp}\rangle \langle \psi_{\rm sp}| )=0.4483$. The difference between the experimental negativity and the ideal one is mainly due to the influence of decoherence, {demonstrated} by the loss of the purity of the density matrix from $1$ to $0.4646$.

\section{Characterization of the super-cat state}
Due to the non-adiabatic effects and the presence of decoherence, the emergent cat state during the SPT contains three superimposed components{:} one corresponding to the empty field mode, while the other two corresponding to the emergent coherent fields with opposite phases, as illustrated in Fig. \ref{EC}.
The size of this super-cat state is given by

\begin{eqnarray}\label{size}
\mathcal{S}=
\frac{\sum_{s\neq l} d_{sl}^2\sqrt{P_sP_l} }{\sum_{s\neq l} \sqrt{P_sP_l}},
\end{eqnarray}
with $\{s,l\}=\{|0\rangle,|\alpha\rangle,|-\alpha\rangle\}$
and $d_{sl}^2$ indicating the square of the phase-space distance between $s$ and $l$. Here $P_{|0\rangle}$, $P_{|\alpha\rangle}$,
and $P_{|-\alpha\rangle}$ are the populations in $|0\rangle$,
$|\alpha\rangle$, and $|-\alpha\rangle$ of $\rho_{e,e}/{\rm Tr}(\rho_{e,e})$ [or $\rho_{g,g}/{\rm Tr}(\rho_{g,g})$],
respectively. For the three-components cat-like state,
mixing $|0\rangle$, $|\alpha\rangle$, and $|-\alpha\rangle$,
Eq. (\ref{size}) becomes

\begin{align}
\mathcal{S}=\frac{\sqrt{P_{|0\rangle}P_{|\alpha\rangle}}|\alpha^2|+\sqrt{P_{|0\rangle}P_{|-\alpha\rangle}}|\alpha^2|+\sqrt{P_{|\alpha\rangle}P_{|-\alpha\rangle}}|2\alpha^2|}{\sqrt{P_{|0\rangle}P_{|\alpha\rangle}}+\sqrt{P_{|0\rangle}P_{|-\alpha\rangle}}+\sqrt{P_{|\alpha\rangle}P_{|-\alpha\rangle}}},
\end{align} 
specifically, $\mathcal{S}_{e,e}=14.03$ and $\mathcal{S}_{g,g}=13.27$ for $\rho_{e,e}/{\rm Tr}(\rho_{e,e})$ and $\rho_{g,g}/{\rm Tr}(\rho_{g,g})$, respectively.

The NP-SP quantum coherences associated with the qubit states $|e\rangle$ and $|g\rangle$ are

\begin{eqnarray}
\mathcal{C}_{e,e}=\sum_{n=1}^{\infty}|\langle 0|\rho_{e,e}|n\rangle|/{\rm Tr}(\rho_{e,e})=1.018,\ \ \ \ 
{\rm and} \ \ \  \
\mathcal{C}_{g,g}=\sum_{n=1}^{\infty}|\langle 0|\rho_{g,g}|n\rangle|/{\rm Tr}(\rho_{g,g})=1.020,
\end{eqnarray}
respectively.
The quantum coherence averaged over these two super-cat states is 1.019.}

Negative values of the Wigner functions clearly distinguish cat states, which are macroscopically-distinct coherent superpositions of classical-like states, from their mixtures. These negative values (as shown by the blue regions in Figs.~\ref{simwig} and \ref{expwigdata}, as well as Fig. 3 in the main text), which are clearly seen between the main peaks (as shown on the
left- and right-hand sides of the figures), occur as a result of interference in the phase space~\cite{SchleichBook}.
{In the digital quantum simulation of the deep-strong coupling dynamics reported in Ref. \cite{Langford2017}, a similar nonclassical state was generated by repetitive application of digital $\pi$ pulses interleaved with short Jaynes-Cummings (JC) interaction without the counter-rotating-wave terms, which allows emulation of the long-time Rabi dynamics, but does not lead to the simultaneous realization of the JC and anti-JC interactions necessary for observing the associated SPT.}

\section{Numerical simulation of the Dicke-model SPT}

{Pushing one step further, we theoretically extend our method to the Tavis-Cummings (TC) model involving multiple qubits coupled to a resonator \cite{Tavis1968,Song2019}. 
By longitudinally modulating and transversely driving each qubit, the TC model can be effectively transformed to the Dicke model with similar parameters.  
The results show that the qubits-resonator system can be evolved from the NP to the SP featuring a highly entangled cat state, formed by two photonic coherent states with opposite phases that are nonclassically correlated with distinct multiqubit coherent states also with opposite phases.} 

The Dicke model, composed of $N$ qubits coupled to a quantum photonic field
mode, is described by the Hamiltonian%
\begin{eqnarray}
H_{D} &=&\delta a^{\dagger }a+\sum_{j=1}^{N}[\frac{\Omega }{2}(\left\vert
e_{j}\right\rangle \left\langle e_{j}\right\vert -\left\vert
g_{j}\right\rangle \left\langle g_{j}\right\vert ) \cr\cr
&&+\frac{\eta }{\sqrt{N}}(\left\vert e_{j}\right\rangle \left\langle
g_{j}\right\vert +\left\vert g_{j}\right\rangle \left\langle
e_{j}\right\vert )(a^{\dagger }+a)],
\end{eqnarray}%
where $\eta $\ denotes the collective qubit-field coupling strength, and $\left\vert g_{j}\right\rangle $ and $\left\vert e_{j}\right\rangle $
denoting the ground and excited states of the $j$th qubit, respectively. In
the thermal limit $N\rightarrow \infty $, the system undergoes a SPT at the
critical point $\xi =2\eta /\sqrt{\Omega \delta }=1$ \cite{Dicke_SPT}. When $\xi $ is
sufficiently large the system has two degenerate ground states, given by
\cite{cat_form}
\begin{eqnarray}
\left\vert \psi _{sp}^{\pm }\right\rangle \simeq \frac{1}{\sqrt{2}}%
[\left\vert \alpha' \right\rangle \prod\limits_{j=1}^{N}\left\vert
+_{j}\right\rangle \pm \left\vert -\alpha' \right\rangle
\prod\limits_{j=1}^{N}\left\vert -_{j}\right\rangle ],
\end{eqnarray}
where $\left\vert \pm _{j}\right\rangle =(\left\vert g_{j}\right\rangle \pm
\left\vert e_{j}\right\rangle )/\sqrt{2}$, and $\left\vert \pm \alpha'
\right\rangle $ represent the coherent states of the photonic field, with $%
\alpha' =\frac{\sqrt{N}\lambda }{\delta }\sqrt{1-\xi ^{-2}}$. For the
even-parity ground state $\left\vert \psi _{sp}^{+}\right\rangle $, the
field parts associated with the even and odd collective excitation numbers
of the qubits are even and odd cat states, $\left\vert C_{\pm }\right\rangle
=(\left\vert \alpha' \right\rangle \pm \left\vert -\alpha' \right\rangle )/\sqrt{%
2}$, respectively.

\begin{figure}[h]
\includegraphics[width=16cm]{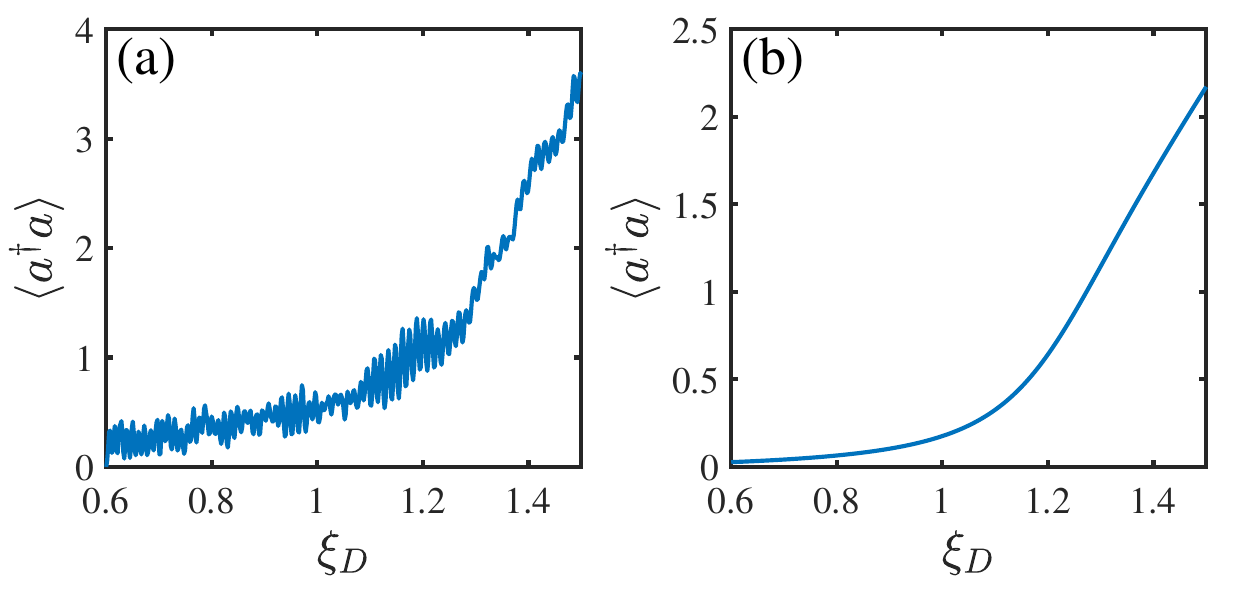}
\caption{Numerically simulated average photon number. (a) Result governed by full Hamiltonian. 
	(b) Result deduced by even ground state of the Dicke-model Hamiltonian.
}
\label{n_compare}
\end{figure}

\begin{figure}[h]
\includegraphics[width=18cm]{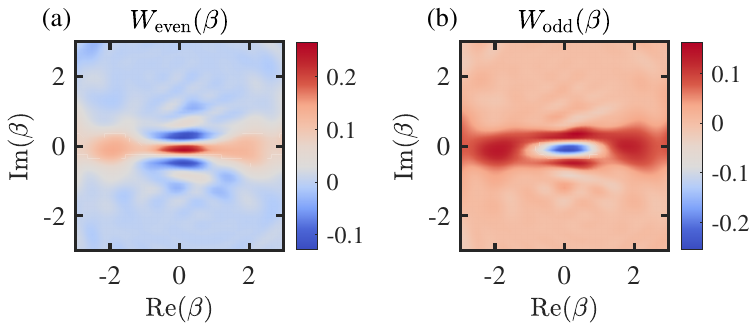}
\caption{Numerical Wigner tomography for final state after the Dicke-model SPT, governed by the full Hamiltonian. 
(a) Wigner functions associated with the
collective even-parities of the qubits.
(b) Wigner functions associated with the
collective odd-parities of the qubits.}
\label{Dicke_SPT_wig_full}
\end{figure}

\begin{figure}[h!]
\includegraphics[width=18cm]{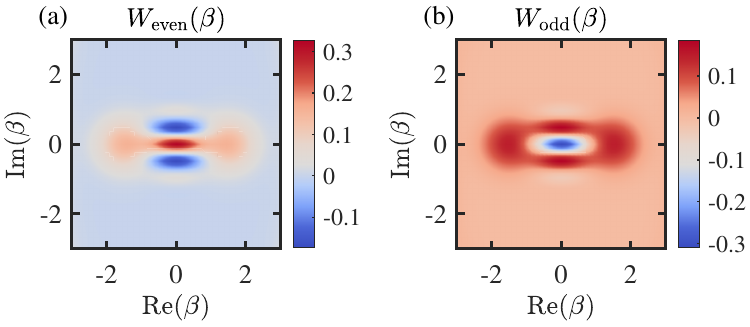}
\caption{Numerical Wigner tomography for final state after the Dicke-model SPT, deduced by the ground state of the Dike-model Hamiltonian. 
(a) Wigner functions associated with the
collective even-parities of the qubits.
(b) Wigner functions associated with the
collective odd-parities of the qubits.}
\label{Dicke_SPT_wig_eig}
\end{figure}

When the coupling strength $\lambda $ between each qubit and the resonator
is much smaller than the qubits' frequency $\omega _{0}$ and the field
frequency $\omega _{p}$, the counter-rotating$-$wave terms for realizing the
Dicke SPT can be effectively realized by applying a resonant transverse
driving with the amplitude $K$, and two longitudinal modulations with
frequencies $\nu _{1}$ and $\nu _{2}$ and amplitudes $\varepsilon _{1}$ and $%
\varepsilon _{2}$, to each of the qubit. Under the conditions $\lambda
,K,\delta =\omega _{p}-\omega _{0}-2\nu _{1}\ll \nu _{1}$ and $\nu
_{2}=2KJ_{0}(\mu )$ with $\mu =\varepsilon _{1}/\nu _{1}$, the system
dynamics in the interaction picture can be effectively described by the
Dicke Hamiltonian with $\eta =\sqrt{N}\lambda J_{2}(\mu )/2$ and $\Omega
=\varepsilon _{2}/2$.

To confirm the validity of the approximations for deriving the effective Hamiltonian, we perform numerical simulations for the 10-qubit Dicke model, in the symmetric Dicke subspace, without considering decoherence.
We here set $\omega _{p}/(2\pi)=5.581$ GHz, $\lambda/(2\pi)=19.91$ MHz, $K/(2\pi)=19.91$ MHz, $\varepsilon_{1}/(2\pi)=207.31$ MHz, $\nu _{1}/(2\pi)=250$ MHz, and $\nu _{2}/(2\pi)=33.28$ MHz. With this
setting, $\xi$ can be controlled by $\omega_{0}$ and $\varepsilon _{2}$. With this
setting, $\xi $ can be controlled by $\omega _{0}$ and $\varepsilon _{2}$.
The average photon number, simulated for the quenching process where $\delta$ and $\varepsilon_{2}$ are varied as $\delta=\varepsilon_{2}=2\eta/\xi(t)$, with $\xi(t)=0.6+0.9t/t_f$ ($t_f$ the total quench time), is shown in Fig. \ref{n_compare}(a),
which well agrees with the result obtained from the even ground state of the
ideal Hamiltonian (Fig. \ref{n_compare}(b)). The quenching process starts with the state $%
\left\vert 0\right\rangle \prod\limits_{j=1}^{N}\left\vert
g_{j}\right\rangle $. The Wigner functions of the field, associated with the
collective even- and odd-parities of the qubits after a quenching time $t=2$ $\mu$s,
are respectively displayed in Figs. \ref{Dicke_SPT_wig_full}(a) and \ref{Dicke_SPT_wig_full}(b). These field states are in well
agreement with those based on the ideal ground states with the same
parameter $\xi$, displayed in Figs. \ref{Dicke_SPT_wig_eig}(a) and \ref{Dicke_SPT_wig_eig}(b), respectively. These results
imply that the presently demonstrated techniques can be used to realize the
Dicke model and the associated SPT.
